\documentclass[preprint, preprintnumbers,nofootinbib,12pt]{revtex4-2}
\usepackage[english]{babel}
\usepackage{color}
\usepackage{amsmath}
\usepackage{physics}
\usepackage[hidelinks]{hyperref}
\usepackage{diagbox}
\usepackage{multirow}
\usepackage{amsfonts}
\usepackage{amssymb}
\usepackage{latexsym}
\usepackage{graphicx}
\usepackage{adjustbox}
\usepackage{xr}
\usepackage{siunitx}
\usepackage{threeparttable}
\usepackage{braket}
\usepackage[justification=centering]{caption}
\usepackage{float}
\usepackage{caption}
\usepackage{subcaption}
\usepackage{mathtools}
\usepackage{array}
\usepackage{seqsplit}
\usepackage{textcomp}
\usepackage{enumitem}
\usepackage{tabularx}
\usepackage[dvipsnames]{xcolor}
\usepackage{soul}

\usepackage{setspace}
\usepackage{soul}
\usepackage{cancel}
\usepackage[scr=rsfs]{mathalpha}
\usepackage{amsmath,bm}
\newcommand{\al}{\alpha}

\newcommand{\pa}{\partial}

\newcommand{\la}{\lambda}

\newcommand{\Om}{\Omega}

\newcommand{\rar}{\rightarrow}

\begin{document}
\title{Radial power-like potentials: from the Bohr-Sommerfeld $\ell \neq 0$ state energies to the exact ones\\}

\author{J.C.~del~Valle}
\email{delvalle@fisica.unam.mx (corresponding author)}

\affiliation{Instituto de F\'isica, Universidad Nacional Aut\'onoma de M\'exico,
	A. Postal 70-543 C. P. 04510, Ciudad de M\'exico, M\'exico.}

\author{A.V.~Turbiner}
\email{turbiner@nucleares.unam.mx, alexander.turbiner@stonybrook.edu }

\affiliation{Instituto de Ciencias Nucleares, Universidad Nacional Aut\'onoma de M\'exico,
	A. Postal 70-543 C. P. 04510, Ciudad de M\'exico, M\'exico.}

\begin{abstract}
Two new Bohr-Sommerfeld (B-S) type quantization conditions for $d$-dimensional radial potentials are proposed 
for non-singular and singular potentials, respectively. 
The power-like potentials \hbox{$V=\operatorname{sign}(m)\,r^m$} for $m=-3/2,-1,-1/2, 1,3,4,6$ and $V=\log r$ 
(that corresponds to $m=0$) are studied in detail for radial quantum numbers $n_r \in [0,100]$ and for angular momenta $\ell=1,2,3,4,5$ in dimensions $d=2,3$. It is shown that for large $n_r$ the B-S energies $E_{BS}$ always tend to the exact ones. For $m=2$ (the radial harmonic oscillator) and $m=-1$ (the $d$-dimensional analogue of the $3D$ Coulomb potential) the B-S spectrum coincides with the exact one  for the first and second quantization conditions, respectively. 
In general, up to a small parametric shift of the radial quantum number, the analytic formulas for the energy spectrum can be written for arbitrary power-like potential. 
A comparison of $E_{BS}$ with the energies $E^{(L)}_{BS}$, obtained for the radial potentials modified by the Langer correction for $d=3$, are in  good agreement in general: in detail, it was checked for the quartic oscillator $V=r^4$. 
It is demonstrated that introducing a dimensionless {\it WKB correction} $\gamma$ into the r.h.s. of the proposed B-S type quantization condition allows us to reproduce the exact energies with~$\gamma$ being very small for positive degrees $m \geq 0$ and sufficiently small for non-positive ones $m < 0$ for all $n_r, \ell$ studied.
%Modification of B-S type quantization condition for which BS spectrum is exact for both the $d$-dimensional radial harmonic oscillator and the Coulomb potential is suggested.
\end{abstract}

\maketitle
\newpage	

\section*{Introduction}

As a continuation of three previous papers \cite{BS-1:2021,BS-1-Add:2025,BS-d:2023}, where a detailed analysis of the Bohr-Sommerfeld (B-S) quantization condition for one-dimensional power-like potentials
\[
V(x)=|x|^m\ ,\ m>0 \ ,
\]
and for the $S$-states of the $d$-dimensional radial power-like potentials
\[
      V(r)\,=\, \frac{m}{|m|}\, r^m\ ,\ m>-2\ , \quad\mbox{and}\quad ,\quad V(r)\,=\,\log r \ ,\ m=0\ ,
\]
was carried out, the states with non-zero angular momentum $\ell$ in radial power-like potentials 
are studied in this work.

In the papers \cite{BS-1:2021,BS-d:2023}, both absolute and relative deviations of the B-S energies $E_{BS}$ from the exact ones $E_{exact}$  were studied for the ground state, low-lying excited and the highly-excited $S$ states. It was demonstrated that, even for the lowest-lying excited states, these energies are sufficiently close—not only for the highly excited states, as is usually proclaimed. Naturally, the maximal deviations occurred for the ground state energy case. Following the seminal paper by Kramers \cite{Kramers}, we also proposed a simple modification of the r.h.s. of the B-S quantization condition for radial confining potentials $V(r)$ at $d > 1$ for the case of $S$-states ($\ell=0$) by introducing the specific $d$-dependent Kramers-type correction $d/2$ (which becomes the familiar 1/2 in the one-dimensional case \cite{Kramers}), and additionally a certain correction term denoted as $\gamma$, {defined through}
\begin{equation}
	\frac{1}{\hbar}\int_{r_A}^{r_B}\sqrt{2\mu [E_{exact}-V(r)]}\,dr\ =\
     \pi\left(n_r\ +\ \frac{d}{4} + \gamma(n_r,d) \right)\ ,\qquad n_r \ =\  0,1,2, ...\ ,
\label{1D}
\end{equation}
where $r_{A,B}$ are the turning points, $\mu$ is the mass of particle, $d$ is spatial dimension and $n_r$ is the radial quantum number. By choosing $\gamma$ in such a way as to reproduce the exact energy spectrum, it turned out that for {\it all} studied radial confining potentials $\gamma=\gamma(n_r,d)$ is a numerically-small, dimensionless, and easily numerically-calculable function; we called it the {\it WKB correction}. It was also checked that for various power-like potentials at $d=1$ this correction systematically $|\gamma| \leq 1/2$\,! Furthermore, it was shown in \cite{BS-1-Add:2025} also that $\gamma$ is a function of the square of the shifted quantum number $I=n+1/2$; i.e., $\gamma=\gamma \left(I^2\right)$.

Evidently, for the harmonic oscillator in any dimension $d$, the WKB correction vanishes: $\gamma=0$.
As for  the square well potential with infinite walls it is a constant at $d=1$: 
it reaches its maximal value $\gamma=1/2$. It is not well-known but quite evident that 
for two exactly-solvable problems in $d=1$, Morse and P\"oschl-Teller $(V(x)=-\la(\la+1)\mbox{sech}^2x)$ 
potentials, the integral on the l.h.s. of (\ref{1D}) can be evaluated analytically, both $\gamma$ are constants, 
which are equal to 0 and ${\sqrt{\la(\la+1)}}-\la-1/2$, respectively. As for the highly-excited states 
$n_r \gg 1$, $\gamma$ tends to zero for any $d$, $\gamma(n_r,d) \sim 1/n_r$ as $n_r \rar \infty$. 
The goal of the present paper is to generalize the quantization rule (\ref{1D}) to the states with non-zero 
angular momentum $\ell \neq 0$ in  $d$-dimensional confining radial power-like potentials.

Needless to say, from the early days of quantum mechanics there were numerous attempts to generalize 
the B-S quantization condition (\ref{1D}) to the $d$-dimensional radial potentials $V(r)$. In this work, 
we propose a generalized, B-S quantization type condition of the form
\begin{equation}
	\frac{1}{\hbar}\int_{r_A}^{r_B}\sqrt{E_{BS}-V(r)}\,dr\ =\
     \pi\left(n_r\ +\ \frac{d+2\ell}{4}\right)\ ,\qquad n_r ,\ell\ =\  0,1,2, ...\ ,
\label{radial-L}
\end{equation}
where $r_{B} > r_A \geq 0$ are the radial turning points, $V(r_A, r_B)=E_{BS}$.~ 
\footnote{Note that $r_A$ is not necessarily a turning point. For example, Eq. \eqref{BSradial} 
was calculated for $r_A=0.$} Here,  $n_r$ is 
the radial quantum number, it is equal to the number of radial nodes in the eigenfunction 
of the $d$-dimensional radial Schr\"odinger equation, namely\footnote{Boundary conditions 
are chosen in such a way that 
$\int_{0}^{\infty}|\psi(r)|^2\,r^{d+2\ell-1}\,dr\,<\, \infty\ , 
\quad d=1,2,...\ $ and $|\psi(r)| \neq \infty $ at $r \rar 0$.},
\begin{equation}
	 	-\frac{\hbar^2}{2\mu}\left(\psi''(r)\ +\ \frac{d+2\ell-1}{r}\psi'(r)\right)+\ V(r)\psi(r)\ =\
        E\psi(r)\ , \qquad r\in[0,\infty) 	\ ,
\label{radial_se}
\end{equation}
where, for the sake of simplicity,  we will set $\hbar=2\mu=1$ from now on. Here, $d$ is the space dimension 
and $\ell$ is the angular momentum. Unlike the standard quantization rules, see e.g.~\cite{Karnakov:2012} for discussion, the quantization rule (\ref{radial-L}) contains 
the original potential $V(r)$ as proposed in \cite{Karnakov-Popov:1999}, and
{\it not} an effective potential which includes the centrifugal term. 
We also exclude the appearance of the Langer-type correction \cite{Langer:1937} 
to the original potential in the l.h.s. of (\ref{radial-L}).

Note that for $d=2$, the quantization condition (\ref{radial-L}) was presented in 
\cite{Berry,Brack} by using the so-called Langer transformation. For arbitrary $d$, 
a quantization condition somewhat similar to (\ref{radial-L}) was proposed in 
\cite{Marinov-Popov}, see Appendix therein, Eq.(A.3), but with a different potential-dependent 
r.h.s.: the quantization rule proposed  there, was applied to singular at origin radial potentials with behavior 
$V(r) \sim r^{m}$ at $r \rar 0$ for $-2 < m < 0$ but decaying exponentially 
at large distances $r \rar \infty$, $V(r) \sim \exp(-a r)$. It was shown that this modification 
leads to reasonably accurate results. Following these works 
\cite{Berry,Brack,Marinov-Popov,Karnakov-Popov:1999}, we propose 
that the quantization condition (\ref{radial-L}) holds for general confining, not necessarily singular 
at origin, radial power-like potentials for any angular momentum and in any dimension $d$.

There are three fundamental properties that the proposed quantization condition (\ref{radial-L}) 
 fulfills:

\begin{itemize}

\item[(i)] At $d=1$ it coincides with the celebrated one-dimensional WKB quantization
     condition\footnote{With  quantum number $N=2\,n_r$.}, if it is assumed an even potential, 
     $V(x)=V(-x)$, with symmetric turning points $x_A=-x_B$\,;
%     while for arbitrary $d$ and $\ell=0$ it corresponds to (\ref{1D}) at $\gamma=0$\,,

\item[(ii)] For the harmonic oscillator potential, $V(r)=r^2$,
      the condition (\ref{radial-L}) leads to the exact energies for any $d$ with the correct multiplicity\,;

\item[(iii)] At large $n_r$ but fixed $d$ and $\ell$, the Bohr-Sommerfeld energies 
$E_{BS}$ from (\ref{radial-L}) approach  the exact ones.
\end{itemize}
	
Property (i) is evident, property (ii) can be easily checked by direct calculation, and property (iii) 
was already verified for $\ell=0$ in \cite{BS-d:2023}. Verification of (iii) for non-vanishing $\ell \neq 0$ 
requires  extensive numerical calculations, which are carried out in this work. 

It must be emphasized that for singular potentials the r.h.s. in (\ref{radial-L}) has to be modified: 
\begin{itemize}
\item[(iv)] in order to provide the exact spectra of the $d$-dimensional Coulomb problem with potential 
$V=-1/r$, it should be 
\begin{equation}
	\frac{1}{\hbar}\int_{r_A}^{r_B}\sqrt{E_{BS}-V(r)}\,dr\ =\
     \pi\left(n_r\ +\ \frac{d+2\ell-1}{2}\right)\ ,\qquad n_r,\ell \ =\  0,1,2, ...\ ,
\label{radial-L-singular}
\end{equation}
for $d > 1$.
This quantization condition will be applied to the potentials with singular behavior at origin
\[
      V\ \rar\ -r^m\ ,\ 0 > m > -2\ \ at \ \ r \rar 0\ .
\]
Thus, we may replace condition (ii) by (iv). 

\end{itemize}

Naturally, the WKB correction $\gamma$, see (\ref{1D}), can be introduced into the r.h.s. of (\ref{radial-L}),
\begin{equation}
	\frac{1}{\hbar}\int_{r_A}^{r_B}\sqrt{E-V(r)}\,dr\ =\ \pi\left(n_r\ +\ \frac{d+2\ell}{4}\ +\
       \gamma_+\right)\ ,
\label{radial_modified}
\end{equation}
for non-singular potentials, or into the r.h.s. of (\ref{radial-L-singular}) for singular ones,
\begin{equation}
	\frac{1}{\hbar}\int_{r_A}^{r_B}\sqrt{E-V(r)}\,dr\ =\ \pi\left(n_r\ +\ \frac{d+2\ell-1}{2}\ +\
       \gamma_-\right)\ ,
\label{radial_modified-singular}
\end{equation}
with the aim to find such a $\gamma_{\pm}$ for which the $E_{BS}$ energies coincide with the exact energies 
$E_{exact}$. It will be checked that in many occasions $|\gamma_{\pm}| \leq 1/2$. 
Certainly, the presence of $\gamma_{\pm}$ in (\ref{radial_modified}) or 
(\ref{radial_modified-singular}) does not break the fundamental properties (i)-(iv). It is worth noting that
explicit indications that the quantization conditions 
can be different for non-singular and singular potentials were already presented in \cite{Karnakov:2012}. 

The formulas (\ref{radial_modified}) and (\ref{radial_modified-singular}) will be the main objects 
of study in the present paper. We will continue to call these formulas (\ref{radial_modified}) and (\ref{radial_modified-singular}) the {\it exact WKB (radial) quantization conditions}. For the potential 
$V=r^4$, it will be checked (and confirmed) that $\gamma_+$ in (\ref{radial_modified-singular}) depends on the square of the r.h.s.:  $$\gamma_+=\gamma_+\bigg(\left(n_r\ +\ \frac{d+2\ell}{4}\right)^2\bigg)\ .$$

{Specifically,} the general radial power-like potential of the form
\begin{equation}
		V(r)\ =\ g^{m-2}\,\mbox{sign}(m)\,r^m\ ,\ m > -2\ ,\ \mbox{sign}(m)=\frac{m}{|m|}\ ,
\label{rpot}
\end{equation}
will be studied, where $g$ is a coupling constant. Signum function is defined as
\begin{equation}
   \mbox{sign}(m)=1\ \mbox{for}\ m \geq 0 \ \mbox{and}\ \mbox{sign}(m)=-1 \ 
   \mbox{for}\ m < 0 \ ,
\label{signum}
\end{equation}
which guarantees the existence of the discrete spectra in (\ref{rpot}).
Let us emphasize that the quantity
\begin{equation}
   \la= \bigg(\frac{\hbar^2}{2\mu}\bigg)^{1/4} g\ , 
\label{lambda}
\end{equation}
is dimensionless and thus the dimensionless WKB correction $\gamma$ can be a function
of $\la$.\footnote{In the present consideration, $r$ is a scaled coordinate satisfying $[r]^2 = [E] = [\hbar]$.}

For the potentials (\ref{rpot}) with positive $m>0$, the integral in the l.h.s. of (\ref{radial-L}) 
and (\ref{radial_modified}) is reduced to the Euler Beta function $B(a,b)$ and the energies can be written 
in closed analytic form. The case $m \leq 0$ requires special attention and it will be discussed 
in Section III, in particular, for the cases of logarithmic and Coulomb potentials.  We will follow 
the same program as in our previous studies \cite{BS-1:2021,BS-d:2023}: first we will calculate and 
compare the energies emerging from the B-S quantization condition (\ref{radial-L}) with the ones obtained 
in highly-accurate numerical calculations of equation (\ref{radial_se}), and then we calculate the absolute 
A.D. and relative R.D. deviations of the B-S from the exact energies. Second, we will calculate 
the WKB correction $\gamma$ in (\ref{radial_modified}) for different power-like potentials 
in different dimensions $d$.

The structure of this paper is the following. In Section I, the quantization rules (\ref{radial-L}) and
(\ref{radial_modified}) are hinted.
Section II is devoted to studying the analytical and numerical properties of $E_{BS}$
for the non-singular power-like potentials (\ref{rpot}) with $m>0$. The explicit calculations
of the WKB correction $\gamma$ for two physically relevant dimensions, $d=2,3$ are presented.
In Section III, we apply the quantization rule (\ref{radial-L}) to two singular potentials 
(\ref{rpot}) with $m \leq 0$: the Coulomb and logarithmic potentials, and two other singular 
potentials, $(-r^{-1/2})$, $(-r^{-3/2})$, and also funnel-type potential, {$V=-Z/r + g\,r$, } 
are mentioned briefly. We explore the accuracy 
of $E_{BS}$ and compute the WKB correction $\gamma_{\pm}$. A modification of the quantization 
rule (\ref{radial-L}), leading to the exact energy spectrum of the Coulomb problem in $d$ 
dimension (\ref{radial_modified-singular}), is proposed. Section IV is dedicated to the comparison of (\ref{radial-L}) 
with the formalism, where the original radial potential is modified by the Langer correction; 
the quartic oscillator potential $r^4$ is taken as the basic example.
The results are summarized in Conclusions. Extensive numerical results are presented in the Appendix.

\section{From S-states to $\ell \neq 0$ states}
\label{extension}

Let us write the original $d$-dimensional Schr\"odinger equation in the Euler coordinates 
$(r, \Om)$, where $r$ is the radius, $r=\sqrt{x_1^2+x_2^2+ \ \dots\ +x_d^2}$, and 
$\{ \Om \}$ is the set of $d-1$ Euler angles. Separating out 
the angular variables, we arrive at the standard \textit{radial} Schr\"odinger equation
\begin{equation}
	-\left(\psi''(r)\ +\ \frac{d-1}{r}\psi'(r)\right)\ +\ \frac{\ell(\ell+d-2)}{r^2}\psi(r)\ +\
                    V(r) \psi(r)\ \ =\ E \psi(r)\ ,
\label{radial_sel}
\end{equation}
where $\ell=0,1,...\,$. The centrifugal term (singular at $r=0$) in the l.h.s. of (\ref{radial_sel}) 
can be removed by taking the radial wave function in the form
\begin{equation}
   \psi(r)\ =\ r^{\ell}\,R(r)\ ,
\end{equation}
which is equivalent to making a gauge rotation of the radial Hamiltonian operator~\footnote{This Hamiltonian 
can be extracted from (\ref{radial_sel}); it reads $\hat{H} \equiv -\left(\pa_r^2+\frac{d-1}{r}\pa_r\right)+\frac{\ell(\ell+d-2)}{r^2}+V(r)$.} with $r^{\ell}$ as the gauge 
factor. As a result, we find that $R(r)$ satisfies a radial Schr\"odinger-type equation,
\begin{equation}
 	-\left(R''(r)\ +\ \frac{d+2\ell-1}{r}\,R'(r)\right)\ +\ V(r)\,R(r)\ =\ E\,R(r)\ .
\label{radial_se-new}
\end{equation}
At $\ell=0$, this equation becomes the one used as the basis for writing the B-S quantization 
condition (\ref{1D}), see \cite{BS-d:2023}, for the $S$ states. This observation suggests 
that the formal replacement
\begin{equation}
      d \rar d+2\ell
\label{rep}
\end{equation}
in (\ref{1D}) should lead to the quantization rule for $\ell$-states (\ref{radial-L}),
\begin{equation*}
	\int_{r_A}^{r_B}\sqrt{E_{BS}-V(r)}\,dr\ =\ \pi\left(n_r\ +\ \frac{d+2\ell}{4}\right)\ .
\end{equation*}
One can easily check that the fundamental properties (i)-(ii) are satisfied. In most of the remaining 
part of the present work, we essentially focus on checking the correctness of the third property (iii) 
by using the same power-like potentials (\ref{rpot}) which were explored in our previous work~\cite{BS-d:2023}. 
Naturally, the WKB correction $\gamma$ can be introduced into (\ref{radial-L}),
\begin{equation*}
		\int_{r_A}^{r_B}\sqrt{E_{exact}-V(r)}\,dr\ =\ \pi\left(n_r\ +\ \frac{d+2\ell}{4}+\gamma_+\right)\ ,
\end{equation*}
see (\ref{radial_modified}), where, in general, $\gamma_+=\gamma_+(n_r,\ell,d)$. We should emphasize that 
by generalizing the argument of H.~Kramers \cite{Kramers} for one-dimensional potentials to 
the effective (radial) potential $V(r)+V_{centrifugal}$, one can show that the $(d+2\ell)/4$ 
term in r.h.s. of (\ref{radial-L}) comes from the linearization of the effective potential near 
the radial turning points $r_{A,B}$ and asymptotic matching, while the corrections to the linear 
term (quadratic, cubic etc) may lead to the appearance of $\gamma_+$.

\section{Non-Singular Power-Like Potentials}
\label{NS}

Let us consider the radial power-like potential
\begin{equation}
	V(r)\ =\ r^m\ ,\qquad m > 0\ ,
\label{rpot-positive_m}
\end{equation}
cf. (\ref{rpot}). Note that this potential (\ref{rpot-positive_m}) coincides with the Green function for \hbox{$(2-m)$}-dimensional Newtonian gravity, being the potential of a point-like source of unit mass 
(up to a normalization constant). The quantization condition (\ref{radial-L}) for potential 
(\ref{rpot-positive_m}) can be evaluated analytically, leading to B-S energies in  closed form
\begin{equation}
	E_{BS}^{(m,d)}\ =\ \left(2 M B
    \left(\tfrac{1}{2},M\right)\left(n_r+\frac{d+2\ell}{4}\right)\right)^{\frac{1}{M}}\ ,
          \qquad M\ =\ \frac{1}{m}\ + \ \frac{1}{2}\ .
	\label{BSradial}
\end{equation}
Here $B(a,b)$ is the Euler beta function, see e.g. \cite{NIST:DLMF}. This generalizes
the analytic expression for the spectrum of one-dimensional power-like potentials obtained in \cite{BS-1:2021}.
Let us emphasize that the $E_{BS}$ energies may be easily analytically continued: from integer 
$n_r, \ell, d$ to complex $n_r, \ell, d$, respectively.
It is evident that $E_{BS}^{(m=2,d)}$ (\ref{BSradial}) lead to the exact energy spectrum \cite{Exact}
of the radial harmonic oscillator $m=2$ for any $d$:
\begin{equation}
	E_{BS}^{(m=2,d)} \ =\  4n_r+d+2\ell\ =\ E_{exact}^{(m=2,d)}\ .
\label{E-m=2}
\end{equation}	

As discussed in \cite{BS-1:2021,BS-d:2023}, the $m=\infty$ case provides an insight into the accuracy 
of the predicted B-S energies. In this limit, the power-like potential describes a radial square well 
of unit length defined in $r\in[0,1]$. In this case, the B-S energy displayed in (\ref{BSradial})  
becomes
\begin{equation}
          E_{BS}^{(m=\infty,d)}\ =\ \pi^2\left(n_r\ +\ \frac{d+2\ell}{4} \right)^2\ .
\label{mInf_approx}
\end{equation}	
On the other hand, for $d>1$, the exact energies obey the transcendental equation
\begin{equation}
\label{Bessel}
	J_{\frac{d+2l-2}{2}}\left(\sqrt{E_{exact}^{m=\infty,d>1}}\right)\ =\ 0\ ,
\end{equation}
where $J_{\nu}$ is the Bessel function of the first kind of order $\nu$, see e.g. \cite{NIST:DLMF}.
In general, equation (\ref{Bessel}) can  be solved  numerically in order to find the energy spectra: 
the only exception appears at $d=3$ where at $\ell=0$ it can be solved analytically  \cite{BS-d:2023}.
Using the asymptotic expansion of the zeros of the Bessel functions; see, e.g., \cite{NIST:DLMF}, keeping
$d$ and $\ell$ fixed, one can establish the asymptotic behavior of the exact energy at large $n_r$,
\begin{equation}
   E_{exact}^{(m=\infty,d)}\ =\ \pi^2\left(n_r+\frac{d+2\ell+1}{4}\right)^2\ +\ \mathcal{O}(n_r^{-1})\ .
\label{mInf_exact}
\end{equation}
cf.(\ref{mInf_approx}).
Consequently, by comparing (\ref{mInf_approx}) and (\ref{mInf_exact}), we can immediately check that 
the condition (iii) is fulfilled for the square-well radial potential. Note that the agreement between 
(\ref{mInf_approx}) and (\ref{mInf_exact}) gets better for larger $(d+2\ell)$. At large $n_r$ 
the WKB correction $$\gamma_+=1/4\ .$$

In order to find the accuracy of the BS energies for finite $m$, we need to compare them with 
the exact energies obtained via a solution of the radial Schr\"odinger equation, which nowadays 
can be solved numerically to any desired accuracy. Based on the experience gained in \cite{BS-1:2021,BS-d:2023}, 
we chose the Lagrange Mesh Method for the numerical calculations  \cite{Mathematica}, which provided 
highly accurate results for the energies with minimal computational effort.  In general terms, 
the Lagrange Mesh Method is an approximate variational approach based on the discretization of 
the time-independent Schr\"odinger equation onto a non-uniform lattice. Further details as well 
as generalities can be found in \cite{Baye}.

\subsection{Numerical Results}

In this section, for angular quantum numbers $1 \leq \ell \leq 5$, we will compare the $E_{BS}$ energies 
with the exact ones $E_{exact}$ obtained using the Lagrange Mesh Method. For the power-like potentials, 
calculations with 1000 mesh points were fast and sufficient to establish the energies of the first 
100 low-lying states, at least, with 4 significant figures for given $\ell$, $d$, and $m$. This comparison 
is done explicitly for the linear, cubic, quartic, and sextic oscillators ($m = 1, 3, 4, 6$, respectively) 
and for two physically important dimensions: $d=2$ and $d=3$. The cases of the logarithmic $(m=0)$ and the 
Coulomb-type $(m=-1)$ potentials will be considered separately below.

In Tables \ref{firstTable-IV}-\ref{fifthTable-VIII}, the results with $d=2$ are presented, while Tables \ref{firstTabled3-IX}-\ref{lastTabled3-XIII} contain the results with $d=3$; for both dimensions the orbital 
quantum numbers $\ell=1,2,3,4,5$ are studied. Those Tables display the $E_{exact}$ and the $E_{BS}$ energies 
for some radial quantum numbers in the interval $0 \leq n_r \leq 20$. Additionally, we include the absolute 
(A.D.) and relative (R.D.) deviations, which are defined as follows
\begin{equation}
    \text{A.D.}\ =\ |E_{BS}-E_{exact}|\ ,\quad	
    \text{R.D.}\ =\ \left|\frac{E_{BS}-E_{exact}}{E_{exact}}\right|\ .
\label{definitions}
\end{equation}
For all potentials we consider, both A.D. and R.D. deviations decrease with growing $n_r$, if the dimension 
$d$ and the angular momentum $\ell$ are kept fixed. For $n_r=20$, the value of the R.D. is $\lesssim 10^{-3}$. 
This is a clear indication that condition (iii) is fulfilled. For the low-lying states of those potentials, 
the A.D. may take large values regardless of $d$, while the R.D. remains sufficiently small. For example, 
if we consider $\ell \geq 2$, then the absolute deviation varies in domain $1 \lesssim$ A.D. $\lesssim 10$.

\begin{figure}[h]
	\centering
	\includegraphics[width=\textwidth]{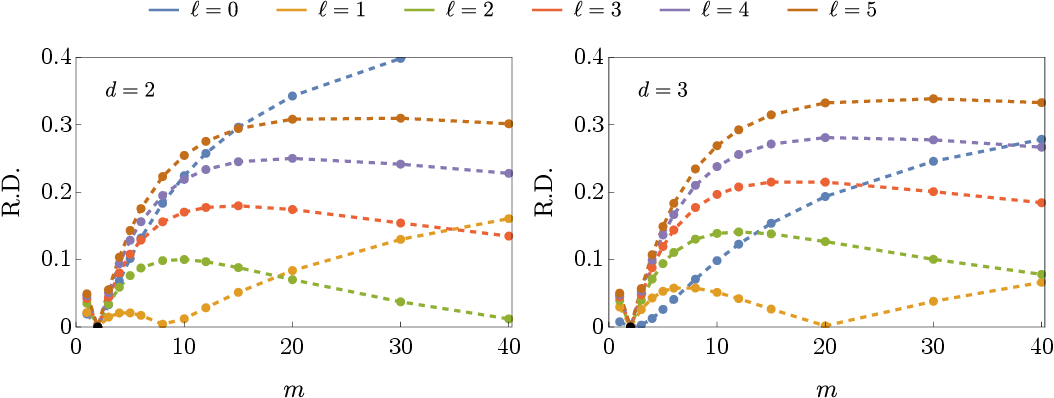}
	\caption {Relative Deviation (R.D.) of the ground state energy, see (\ref{definitions})
             {\it versus} $m$ for angular momenta $\ell=1,2,3,4,5$ in dimensions $d=2,3$.
             Bullets indicate actual calculations; dashed curves drawn for convenience.
             At $m=2$, marked by black bubble, the relative deviation is
             zero.}
	\label{RD}
\end{figure}

In Fig. \ref{RD}, the relative deviation R.D. as a function of $m$ is displayed for the ground state 
($n_r=0$) at $d=2,3$. Surprisingly, in the domain $1 \leq m \leq 40$, the R.D. may develop two minima 
(see e.g. {yellow} curve). 
However, for sufficiently large $m$ and $\ell>3$, the R.D.
decreases monotonically, approaching a finite, nonzero asymptotic limit for both $d=2$ and $d=3$.

\begin{figure}[h]
	\centering
	\includegraphics[width=\textwidth]{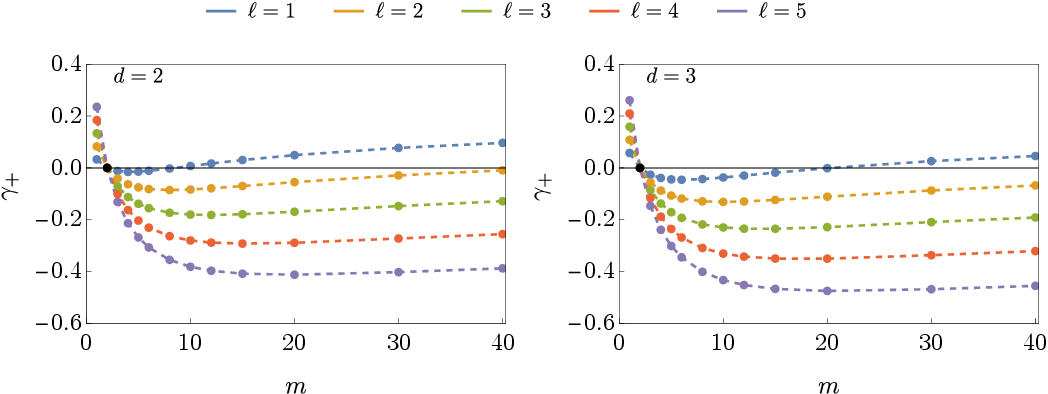}
	\caption {WKB Correction $\gamma_+$ for the ground state energy, see (\ref{gamma})
             {\it versus} $m$ for the angular momenta $\ell=1,2,3,4,5$
             in dimensions $d=2,3$. Bullets indicate actual calculations; dashed curves drawn
             for convenience. At $m=2$, see black bubble, the WKB correction vanishes. }
	\label{WKB_GS}
\end{figure}

\subsection{Exact Quantization Condition}

In the case of radial power-like potentials (\ref{rpot-positive_m}), the exact WKB (radial) quantization 
condition (\ref{radial_modified}) can be solved with respect to the energy, leading to a closed  
expression for the spectrum,
\begin{equation}
	E_{exact}^{(m>0,d)}\ =\ \left(2 M B
    \left(\tfrac{1}{2},M\right)\left(n_r+\frac{d+2\ell}{4} + \gamma_+ \right) \right)^{\frac{1}{M}}\quad ,
          \quad M\ =\ \frac{1}{m}\ + \ \frac{1}{2}\ ,
	\label{BSradial-exact}
\end{equation}
for non-singular at origin potentials, $m > 0 $.
The practical value of this formula is related to the question of the magnitude of $\gamma_+$:
if $\gamma_+$ is relatively small with a negligible dependence on its parameters formula (\ref{BSradial-exact}) 
can provide information on the dependence of the spectrum on $d, n_r, \ell$.
It is still unclear how to calculate $\gamma_+$ analytically/theoretically. Thus, we will proceed to 
this question numerically.

\subsection{WKB Correction}

Once the exact energies are known, the WKB correction $\gamma$ can be unambiguously found according to (\ref{radial_modified}) or (\ref{BSradial-exact}),
\begin{equation}
   \gamma_+\ =\ 	
   \frac{(E_{exact})^M-\frac{1}{2}MB(\tfrac{1}{2},M)(4n_r+d+2\ell)}{2MB(\tfrac{1}{2},M)}\ .
\label{gamma}
\end{equation}
The exact energies can be obtained via the Lagrange Mesh Method with practically any desired accuracy: we made accurate calculations of $\gamma$ for the  power-like potentials at $m=1,3,4,6$. Results for $d=2,3$ at $n_r=0,5,10,20$ and $\ell=1-5$ are presented in Tables \ref{firstTable-IV}-\ref{lastTabled3-XIII}.
In Fig. \ref{WKB_GS}, the ground state WKB correction $\gamma_+$ is displayed as a function of $m$ for $\ell=0,1,2,3,4,5$, Fig. 4 therein. According to the calculations, $|\gamma_+|$ is always largest for the ground state $n_r=0$, and then it begins to decrease at a slow pace as $n_r$ increases. Again, this is an indication that condition (iii) is fulfilled. As occurred in the one-dimensional case and for the $S$-states of radial potentials, the WKB correction for the excited states $\gamma$ is, as was guessed earlier, a smooth and small function (at least, in comparison with the quantum number $n_r$).
Furthermore, $\gamma_+$ decays like $n_r^{-1}$ for large $n_r$.

In principle, one needs to know the whole spectrum of a given potential to find $\gamma_+$ for all values of $n_r$. Conversely, at present one needs to know $\gamma_+$ to find $E_{exact}$. At  first glance, this may look as unattainable.  However, in practice one can calculate the first low-lying energies and then use a fitting function to find $\gamma_+$ up to some accuracy. This program was successfully realized in our previous studies \cite{BS-1:2021,BS-d:2023}. In particular, it was observed that fitting the numerical values of $\gamma_+$ is much easier than fitting the energy as a function of $n_r$ directly. Below, we show that the same observation holds for the $\ell \neq 0$ states.

\begin{figure}[h!]
	\centering
	\includegraphics[width=\textwidth]{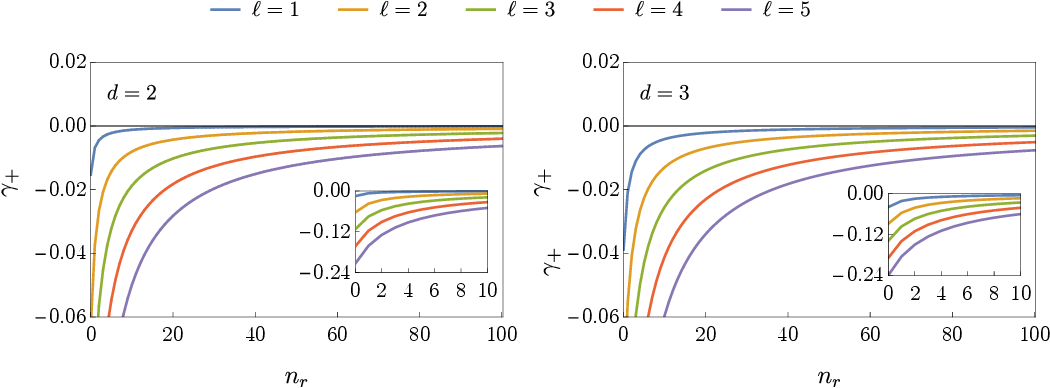}
	\caption {WKB Correction $\gamma_+$ for the quartic potential $V(r)=r^4$ {\it vs} $n_r$
    for $\ell=1,2,3,4,5$ and $d=2,3$. }
	\label{WKB_Quartic}
\end{figure}

As illustration, let us consider the quartic oscillator with potential
\begin{equation}
	V(r)\ =\ r^4 .
	\label{quartic}
\end{equation}			
In Fig. \ref{WKB_Quartic}, the WKB correction $\gamma_+$ {\it vs} $n_r$ is shown in the domain $0 \leq n_r \leq 100$ at $d=2,3$ for all values of $\ell=1,2,3,4,5$.
Note that for $d$ and $\ell$ fixed, the WKB correction can be easily fitted by
\begin{equation}
    \gamma_{fit}(n_r)\ =\ -\frac{a}{b\,n_r+1}\ ,
\label{quarticWKB}
\end{equation}
with sufficiently high accuracy, where $a$ and $b$ are free parameters, which usually demonstrate a  
smooth dependence on $d$ and $\ell$. They are found via least-squares optimization. The fitting function 
(\ref{quarticWKB}) is a simplified version of the function used in \cite{BS-d:2023} to study the $S$-states, 
see Eqs. (20)--(21) therein. Note that at large $n_r$, the fit (\ref{quarticWKB}) leads to the correct asymptotic 
behavior~$\sim n_r^{-1}$  for $\gamma_+$. In Table \ref{parameters},  the optimal values of $a,b$ are shown 
for the angular momentum $\ell=1,2,3,4,5$ at $d=2$ and $d=3$.  With the optimal parameters from Table 
\ref{parameters}, the fitting function $\gamma_{fit}$ (\ref{quarticWKB}) predicts all the energies with 
around 4 exact significant digits if substituted into (\ref{BSradial-exact}). Needless to say that a similar 
fitting function can be used in studies of the WKB correction for other power-like potentials. If a higher 
accuracy is desired, the fitting function is modified
\begin{equation}
	\gamma_{fit}(n_r)\ =\ \frac{P_L(n_r)}{Q_{L+1}(n_r)}
\label{fit-rational}
\end{equation}
where $P_L$ and $Q_{L+1}$ are polynomials of degrees $L$ and $(L+1)$, respectively.
\begin{table}[h]
	\caption{Quartic oscillator $(m=4)$. Parameters $a$ and $b$ for the fitting function (
     \ref{quarticWKB}) at $d=2,3$ and $\ell=1,...,5$.}
	{\setlength{\tabcolsep}{0.4cm}		\begin{tabular}{cccccc}
\hline
\hline	
		& \multicolumn{5}{c}{	\rule{-2pt}{4ex}		$d=2$} \\
\hline
			\rule{-2pt}{4ex}
		$\ell$ & 1   & 2   & 3   & 4  & 5  \\[10pt]
\hline	
			\rule{-2pt}{4ex}		    	
		$a$     & 0.0152  & 0.0630  & 0.1126  & 0.1629  & 0.2137  \\
		$b$     & 1.1887  & 0.6816  & 0.4990  & 0.3953  & 0.3277  \\[5pt]
\hline
\hline
		& \multicolumn{5}{c}{	\rule{-2pt}{4ex}$d=3$} \\	
\hline
			\rule{-2pt}{4ex}	
		$\ell$   & 1   & 2   & 3   & 4  & 5  \\[10pt]
			    \hline	
	\rule{-2pt}{4ex}				
		$a$     & 0.0388  & 0.0877   & 0.1377  & 0.1883   & 0.2392  \\
		$b$     & 0.8446  & 0.5754   & 0.4410  & 0.3583   & 0.3020  \\[5pt]
		\hline
		\hline
	\end{tabular}}
\label{parameters}
\end{table}

Analyzing the data in Tables \ref{firstTable-IV}-\ref{lastTabled3-XIII} one can see that 
the WKB correction $\gamma_+$ is {\it always} a small  function for power-like radial 
potentials (\ref{rpot-positive_m}) with $m = 1,3,4,6$,
\[
     |\gamma_+|\ < \ \frac{1}{2}\ ,
\]
for $d=2,3$ and any $n_r \geq 0$ and angular momentum $\ell=1,2,3,4,5$. This is the same behavior 
as in the one-dimensional case \cite{BS-1:2021} for any positive degree $m>0$ and
for the $S$-states for $d=2,3,6$ of radial power-like potentials \cite{BS-d:2023} with $m = 1,3,4,6$.
One can guess that for any non-singular power-like radial potential the WKB correction $\gamma$ is small 
for any $d \geq 1$, any $n_r \geq 0$ and any angular momentum $\ell$ (for $d \geq 2$), it obeys the inequality
\begin{equation}
\label{inequality}
  |\gamma_+|\ \leq \ \frac{1}{2}\ .  
\end{equation}

\section{Singular Power-Like Potentials}
\label{SP}
\subsection{Coulomb Potential}

If $m=-1$ in (\ref{rpot-positive_m}) we arrive at
\begin{equation}
	V(r)\ =\ -\frac{1}{r}\ ,
\label{m=-1}
\end{equation}	
which corresponds to the {\it celebrated} Coulomb potential at $d=3$, where $r=\sqrt{x^2+y^2+z^2}$. 
It is well known that for arbitrary dimension $d>1$, the energy spectrum for potential (\ref{m=-1}) 
can be found explicitly,
\begin{equation}
	E^{(m=-1)}_{exact}\ =\  -\frac{1}{(2n_r+d+2\ell-1)^2}\ .
\label{Exact_Coulomb}
\end{equation}	
In particular, for $d=3$, the spectrum reads
\[
   E^{(m=-1)}_{exact}(d=3)\ =\  -\frac{1}{4 n^2}\ ,\
\]
see e.g. \cite{LL:1977}, where $n=n_r+\ell+1, n=1,2,3,\ldots$ is the principal quantum number. 
Note that for given $d$ the degeneracy of the spectrum is equal to $n$, if the degeneracy 
with respect to magnetic quantum number(s) is not taken into account.

In turn, the quantization rule (\ref{radial-L}) results in
\begin{equation}
	E_{BS}\ =\ -\frac{1}{\left(2n_r + \dfrac{d}{2} + \ell \right)^2}\ ,
\label{BS_Coulomb}
\end{equation}	
while the exact WKB quantization condition (\ref{radial_modified}) leads to
\begin{equation}
	E\ =\ -\frac{1}{\left(2n_r + \dfrac{d}{2} + \ell + 2 \gamma_+ \right)^2}\ .
\label{BS_Coulomb-exact}
\end{equation}	
It is evident that the $E_{BS}$ and $E_{exact}$ are close asymptotically when $n_r$ is large
at fixed $d, \ell$. This ensures that (iii) is fulfilled. By comparing (\ref{Exact_Coulomb}), 
(\ref{BS_Coulomb}) and (\ref{BS_Coulomb-exact}) one can find the value of the WKB correction,
\begin{equation}
	\gamma_+\ =\ \frac{1}{4}\left(d+2\ell-2\right)\ ,
\end{equation}
which has no $n_r$-dependence. Note that for fixed $d$, the WKB correction $\gamma_+$ grows linearly 
with the increase of $\ell$. Furthermore, this WKB correction vanishes for $d=2$ and $\ell=0$ only, 
implying that the energies $E_{BS}$ coincide with the exact ones for the $S$-states in the planar
 problem \cite{BS-d:2023}. The energies $E_{exact}$ and $E_{BS}$, their A.D. and R.D. deviations, 
 and the WKB correction $\gamma_+$ are shown in Tables \ref{firstTable-IV}-\ref{lastTabled3-XIII}.

It is worth noting that a modification of the B-S quantization condition (\ref{radial-L}),
\begin{equation}
	\frac{1}{\hbar}\int_{r_A}^{r_B}\sqrt{E^{(mod)}_{BS}-V(r)}\,dr\ =\
     \pi\left(n_r\ +\ \frac{d+2\ell-1}{2}\right)\ ,\qquad n_r \ =\  0,1,2, ...\ ,
\label{radial-L-mod}
\end{equation}
\noindent
where $r_A=0$, leads to the exact B-S spectrum for the Coulomb potential (\ref{m=-1}), 
$$E^{(mod)}_{BS}=E^{(m=-1)}_{exact}\ ,$$ for any $d$ and $\ell$. It suggests that a certain modification
of (\ref{radial_modified}) by introducing the WKB correction $\gamma_{-}$ into (\ref{radial-L-mod}),
\begin{equation}
	\frac{1}{\hbar}\int_{r_A}^{r_B}\sqrt{E-V(r)}\,dr\ =\
     \pi\left(n_r\ +\ \frac{d+2\ell-1}{2} + \gamma_{-}\right)\ ,\qquad n_r \ =\  0,1,2, ...\ ,
\label{radial-L-mod-nega}
\end{equation}
c.f.(\ref{radial_modified}), might be the right quantization rule to be used for singular 
at $r=0$ power-like potentials for which $\gamma_{-}$ is small and vanishes for the Coulomb problem, 
$\gamma_{-}=0$. This ensures that (iv) is fulfilled. 

One can  {\bf conjecture} that (\ref{radial-L-mod-nega}) would lead to small $\gamma_{-}$ 
for the general, singular at $r=0$ potentials. Taking the Yukawa potential,
\[
      V\ =\ -\frac{e^{-\al r}}{r}\ ,
\]
as an example with $\al=0.595 306$, see e.g. \cite{DBN:2018} , for which the ground state energy 
is equal to zero, $E=0$, one finds that
\[
    \gamma_+\ =\ 0.284 117\ ,\ \gamma_{-}\ =\ 0.034 117\ .
\]
This example supports the conjecture.

\subsection{Exact Quantization Condition for Singular otentials}

In the case of radial power-like potentials (\ref{rpot-positive_m}) for negative $m < 0$ the exact 
WKB (radial) quantization condition (\ref{radial-L-mod-nega}) can be solved with respect 
to the energy, leading to a closed analytic expression for the spectrum,
\begin{equation}
	E_{exact}^{(m<0,d)}\ =\ 
	-\left[
	\frac{2\pi}{B\left(\frac12,-M\right)}
	\left(n_r+\frac{d+2\ell-1}{2}+\gamma_-\right)
	\right]^{1/M}, \quad M=\ \frac{1}{m}\ +\ \frac{1}{2},
	\label{BSradial-exact-nega}
\end{equation}
cf. (\ref{BSradial-exact}), for singular at origin potentials, $m < 0 $.
The practical value of this formula is related to the question of the magnitude of $\gamma_{-}$:
assuming that $\gamma_-$ is relatively small with a negligible dependence on its parameters 
formula (\ref{BSradial-exact-nega}) 
can provide information on the dependence of the spectrum on $d, n_r, \ell$.
It is still unclear how to calculate $\gamma_-$ analytically/theoretically.

\subsection{Logarithmic Potential}

The logarithmic potential appears as a limiting case of the power-like ones with positive degrees 
$m > 0$ by noting that
\begin{equation}
\label{log-pot}
	V(r)\ =\ \log r\ =\ \lim_{m \to +0}\ \frac{1}{m}(r^m\ -\ 1)\ .
\end{equation}
Hence, this potential can be considered formally as a non-singular potential!	

For any dimension $d$ at any $n_r$ and $\ell$, the energy spectrum of the potential (\ref{log-pot}) 
can be found numerically with the Lagrange Mesh Method to any desired accuracy. On the other hand, 
the quantization rule (\ref{radial-L}) leads to a closed analytic expression for the B-S energies, which reads
\begin{equation}
    E_{BS}\ =\  \log \left[2\sqrt{\pi}\left(n_r+\frac{d+2\ell}{4}\right)\right]\ ,
\label{BS-log-potential}
\end{equation}
while the quantization rule (\ref{radial-L-mod}) leads to the alternative B-S energies,
\begin{equation}
    E^{(mod)}_{BS}\ =\  \log \left[2\sqrt{\pi}\left(n_r+\frac{d+2\ell-1}{2}\right)\right]\ ,
\label{BS-log-potential-mod}
\end{equation}
Hence, the formula for the exact energies has either form
\begin{equation}
    E_{exact}\ =\  \log \left[2\sqrt{\pi}\left(n_r+\frac{d+2\ell}{4} + \gamma_+\right)\right]\ ,
\label{log-potential}
\end{equation}
where  $\gamma_+$ is the WKB correction in (\ref{radial_modified}), or
\begin{equation}
    E_{exact}\ =\  \log \left[2\sqrt{\pi}\left(n_r+\frac{d+2\ell-1}{2} + \gamma_{-}\right)\right]\ ,
\label{log-potential-mod}
\end{equation}
with $\gamma_{-}$ as a new WKB correction in (\ref{radial-L-mod-nega}).

Before making a detailed study of B-S quantization rule let us make a check what is the form of the r.h.s. 
is more relevant for the logarithmic potential: the one which appears in (\ref{radial_modified}) or in (\ref{radial-L-mod-nega}). As a test we consider the ground state $(n_r=0, \ell=0)$ in dimension $d=3$. 
In the Lagrange Mesh method this ground state energy is easily calculated,
\[
      E_{(0,0)}\ =\ 1.044 332\ ,
\]
see \cite{BS-d:2023}. It immediately leads to the WKB corrections,
\[
   \gamma_+(0,0)\ =\ 0.051 573\ \ ,\ \ \gamma_{-}(0,0)\ =\ -0.198 428\ ,
\]
in (\ref{radial_modified}) and (\ref{radial-L-mod-nega}), respectively. Smallness of $\gamma_+$ confirms 
the guess that the logarithmic potential can be treated by using the quantization condition (\ref{radial_modified}) 
relevant for the non-singular potentials.

\begin{figure}[h]
	\centering
	\includegraphics[width=\textwidth]{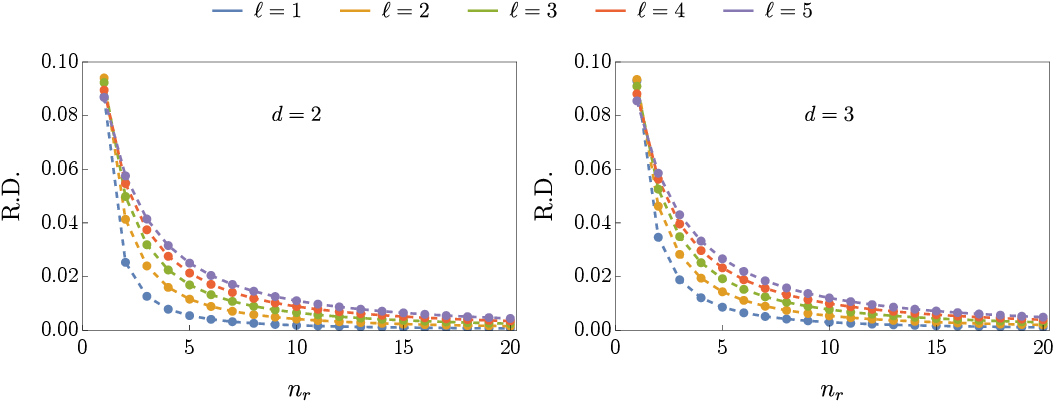}
	\caption {Logarithmic potential: Relative Deviation (R.D), see (\ref{definitions}),
     as a function of  $n_r$ for the angular momenta $\ell=1,2,3,4,5$ and dimension $d=2,3$. Bullets indicate actual calculations; dashed curves drawn for  illustration. }
	\label{RD_LOG}
\end{figure}

The quantitative comparison between the exact $E_{exact}$ and the B-S low-lying $E_{BS}$ energies is found in Tables \ref{firstTable-IV}-\ref{lastTabled3-XIII} for dimensions $d=2,3$ and angular momenta $\ell=1,2,3,4,5$.
%Note that the energy $E_{BS}$ is unexpectedly accurate in comparison with other studied power-like potentials.
According to the calculations, the maximum deviation ($\sim 10\%$) between $E_{exact}$  and $E_{BS}$ occurs for the ground state similarly to all other studied power-like potentials. Then, as $n_r$ increases the R.D. drops rapidly for both $d=2,3$ cases, see Fig. \ref{RD_LOG}.
 \begin{figure}[h]
	\centering
	\includegraphics[width=\textwidth]{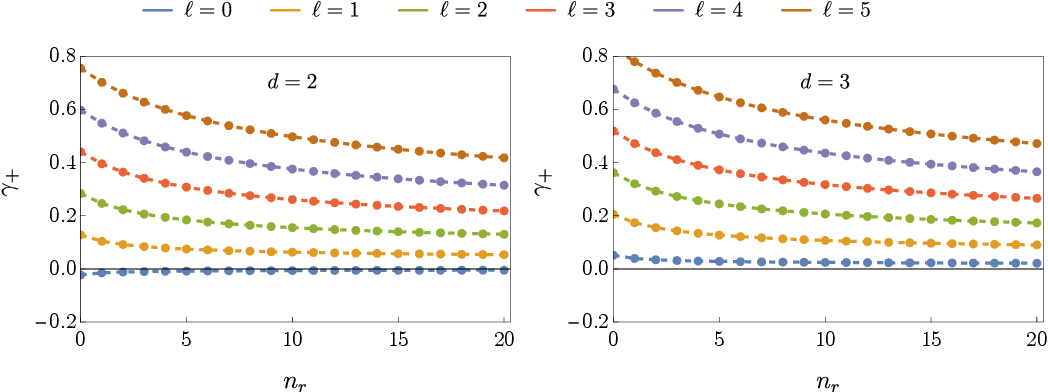}
	\caption {Logarithmic potential: WKB Correction $\gamma_+$, see (\ref{radial_modified}), as
     a function of $n_r$ for the angular momenta $\ell=1,2,3,4,5$ and dimension $d=2,3$. Bullets indicate actual calculations; dashed curves drawn for  illustration, see (\ref{logfit}).}
	\label{WKB_LOG}
\end{figure}

The WKB correction $\gamma_+$ for $d=2,3$ and $\ell=1,2,3,4,5$ demonstrates a smooth decreasing behavior with the increase of $n_r$, while for fixed $n_r$ it grows with the increase of $\ell$, see Fig. \ref{WKB_LOG}. In order to reach four significant digits in energy, the WKB correction $\gamma$ can be fitted by a rational function of the type (\ref{fit-rational}),
\begin{equation}
    \gamma_{fit}^{\log(r)}(n_r)\ =\ \frac{a_1\,n_r\ +\ a_0}{b_2\,n_r^2\ +\ b_1\,n_r\ +\ 1}\ ,
\label{logfit}
\end{equation}
cf.(\ref{quarticWKB}), where the optimal parameters $\{a_0,a_1,b_1,b_2\}$ are shown in Table \ref{parameters_log}. 
This fit is more complicated than (\ref{fit-rational}), which we used in order to get four significant digits 
correctly in the energy of the quartic oscillator, since it involves two extra parameters. However, it is less 
complicated than the one used for the case $\ell=0$ in \cite{BS-d:2023}.
Note that for fixed $d$, all parameters are monotonic functions in $\ell$: $a_{0,1}$ are growing, while 
$b_{1,2}$ are decreasing. It is worth emphasizing that for the fixed $\ell$
the maximal value of the $\gamma_+$ corresponds to $n_r=0$, it is given by $a_0$. It grows with increasing $
\ell$ while being always much smaller than $\ell/2$,
%\newpage
\noindent
see (4)~\footnote{At $\ell=0$ at $d=2$ the $\gamma_+$ is negative and $\max \gamma_{+}$ at $n_r=0$ is equal 
to 0.022, while at $d=3$ the $\gamma_+$ is positive and $\max \gamma_{+}$ is equal to 0.052}.

\begin{table}[h]
	\caption{Logarithmic Potential. Parameters $a$'s and $b$'s for fitting function
             (\ref{logfit}) at $d=2,3$ and $\ell=1,...,5$.}
{\setlength{\tabcolsep}{0.4cm}		
\begin{tabular}{cccccc}
\hline\hline	
			& \multicolumn{5}{c}{	\rule{-2pt}{4ex}		$d=2$} \\
\hline
			\rule{-2pt}{4ex}
			$\ell$ & 1   & 2   & 3   & 4  & 5  \\[10pt]
\hline	
			\rule{-2pt}{4ex}		    	
			$a_0$    & 0.12816   & 0.28383   & 0.44064   & 0.59792  & 0.75526    \\
			$a_1$    & 0.02577   & 0.03683   & 0.04328   & 0.04767  & 0.05091    \\
			$b_1$    & 0.46798   & 0.29643   & 0.21953   & 0.17509  & 0.14595    \\
			$b_2$    & 0.00408   & 0.00228   & 0.00149   & 0.00106  & 0.00080    \\[5pt]
\hline\hline
			& \multicolumn{5}{c}{	\rule{-2pt}{4ex}$d=3$} \\	
\hline
			\rule{-2pt}{4ex}	
			$\ell$   & 1   & 2   & 3   & 4  & 5  \\[10pt]
\hline	
			\rule{-2pt}{4ex}				
			$a_0$     & 0.20577  & 0.36216  & 0.51921   & 0.67654  & 0.83401  \\
$a_1$       &0.03218  & 0.04041  & 0.04566  & 0.04940   & 0.05224    \\
$b_1$       &0.36151  & 0.25198  & 0.19472  & 0.15916   & 0.13482    \\
$b_2$       &0.00297  & 0.00182  & 0.00125  & 0.00091   & 0.00071    \\[5pt]
\hline\hline
\end{tabular}}
\label{parameters_log}
\end{table}

\subsection{Two singular potentials: $(-r^{-1/2})$ and $(-r^{-3/2})$}

{\bf (i)}\ Let us take the singular potential
\begin{equation}
\label{minus1/2}
  V(r)\ =\ -r^{-1/2}\ .
\end{equation}
This potential has an infinite discrete spectrum.
In the Lagrange Mesh method without any difficulties one can calculate the ground state energy
at $d=3$,
\[
      E_{(0,0)}\ =\ -0.438\,041\ ,
\]
which corresponds to
\[
    \gamma_+(0,0) = 0.112\,318\ ,\ \gamma_{-}(0,0)\ =\ -0.137\,682 \ .
\]

{\bf (ii)}\ The same study as in {\bf (i)} can be carried out for the singular potential
\begin{equation}
\label{minus3/2}
  V(r)\ =\ -r^{-3/2}\ .
\end{equation}
In the Lagrange Mesh method one can calculate the ground state energy
at $d=3$,
\[
      E_{(0,0)}\ =\ -0.298\,630\,3\ ,
\]
which corresponds to
\[
    \gamma_+(0,0)\ =\ 0.668\,352\ ,\ \gamma_{-}(0,0)\ =\ 0.418\,352 \ .
\]

\subsection{Funnel-type potential}
\label{sec:funnel}

As a further test of the proposed quantization conditions, we consider the
three-dimensional funnel-type potential
\begin{equation}
	V(r)\ =\ -\frac{Z}{r}\ +\ g\,r\ ,\qquad Z, g>0\ ,
\label{funnel-potential}
\end{equation}
which combines a term with Coulomb singularity at the origin with linear, confinement-type term 
at large distances. Unlike the potentials discussed earlier, this potential involves two competing
terms, singular and non-singular at origin, and therefore provides a useful test ground
for checking the accuracy for quantization conditions \eqref{radial-L} and \eqref{radial-L-mod} 
\footnote{In this subsection, we adopt units $\hbar=\mu=1$}. We use the parameters $Z=0.68812$ and $g=1/2$, 
which were found to be suitable for the description of the bound states of charmonium -  
the quark-antiquark system made from charm quarks $c, \bar c$, see e.g. \cite{Badalyan}.
 
Numerical results are presented in Table~\ref{parameters_funnel}
for selected radial quantum numbers $n_r$ and angular momenta
$\ell=0,5$. Note that the quantization rule \eqref{radial-L} provides the B-S energies, 
which systematically underestimate the exact energies being always below them.
This contrasts with the modified quantization rule \eqref{radial-L-mod} which
systematically overestimates the exact energies being above them. Nevertheless,
for both angular momenta, its relative deviation decreases with increasing $n_r$, 
with the largest deviation occurring for the ground state as expected. 

The accuracy of the two quantization rules can also be assessed
in terms of the WKB corrections $\gamma_{\pm}$.
For the funnel-type potential (\ref{funnel-potential}), the original quantization rule
\eqref{radial-L} generally provides more accurate energies,
particularly for $\ell=5$, where it also yields substantially
small WKB corrections $\gamma_+$ in magnitude; 
see Table~\ref{parameters_funnel}.
In spite of the fact, the modified quantization rule \eqref{radial-L-mod} performs 
better for the lowest radial excited states for $\ell=0$, including the ground state, and 
it leads to very bad behavior of $\gamma_-$ for $\ell=5$. For both angular momenta $\gamma_-$ 
grows with the increase of $n_r$, {which can be considered as an undesirable feature}. 
The overall better performance of the non-modified quantization rule \eqref{radial-L} is consistent
with the dominance of the non-singular linear term when we find
the bound-state energies for the chosen parameters $g$ and $Z$, see above. However, 
$|\gamma_+|$ gets above 1/2 for $\ell=5$ which we set as undesirable property for power-like potentials.

\begin{table}[h]
	\caption{Funnel-type potential (\ref{funnel-potential}).
		Comparison of exact energies with Bohr--Sommerfeld \eqref{radial-L} and modified
		Bohr--Sommerfeld \eqref{radial-L-mod} predictions for selected radial states with
		$\ell=0$ and $\ell=5$. The absolute (A.D.) and relative (R.D.)
		deviations are calculated with respect to the corresponding
		Bohr--Sommerfeld predictions.}
	{\setlength{\tabcolsep}{0.22cm}
		\small
		\begin{tabular}{cc|cccc|cccc}
			\hline\hline
			& \multicolumn{9}{c}{\rule{-2pt}{4ex}$\ell=0$} \\
			\hline
			\rule{-2pt}{4ex}
			$n_r$ & $E_{exact}$
			& $E_{BS}$ & A.D. & R.D. & $\gamma_{+}$
			& $E_{BS}^{(mod)}$ & A.D. & R.D. & $\gamma_{-}$ \\[10pt]
			\hline
			\rule{-2pt}{4ex}
			0   & 0.4902  & 0.0994  & 0.3908 & 0.7974 & 0.2092 & 0.5544  & 0.0642 & 0.1310 & $-0.0408$ \\
			5   & 4.2799  & 4.2017  & 0.0782 & 0.0183 & 0.1425 & 4.3385  & 0.0586 & 0.0137 & $-0.1075$ \\
			10  & 6.6788  & 6.6238  & 0.0551 & 0.0082 & 0.1267 & 6.7322  & 0.0534 & 0.0080 & $-0.1233$ \\
			20  & 10.4946 & 10.4561 & 0.0384 & 0.0037 & 0.1115 & 10.5422 & 0.0476 & 0.0045 & $-0.1385$ \\
			50  & 19.1917 & 19.1681 & 0.0236 & 0.0012 & 0.0929 & 19.2315 & 0.0398 & 0.0021 & $-0.1571$ \\
			100 & 30.3796 & 30.3634 & 0.0162 & 0.0005 & 0.0802 & 30.4138 & 0.0342 & 0.0011 & $-0.1698$ \\[5pt]
			\hline\hline
			& \multicolumn{9}{c}{\rule{-2pt}{4ex}$\ell=5$} \\
			\hline
			\rule{-2pt}{4ex}
			$n_r$ & $E_{exact}$
			& $E_{BS}$ & A.D. & R.D. & $\gamma_{+}$
			& $E_{BS}^{(mod)}$ & A.D. & R.D. & $\gamma_{-}$ \\[10pt]
			\hline
			\rule{-2pt}{4ex}
			0   & 3.0775  & 2.6639  & 0.4136 & 0.1343 & 0.6125 & 4.3385  & 1.2610 & 0.4097 & $-2.1375$ \\
			1   & 3.6639  & 3.3249  & 0.3391 & 0.0926 & 0.5585 & 4.8645  & 1.2006 & 0.3277 & $-2.1915$ \\
			5   & 5.6903  & 5.4830  & 0.2073 & 0.0365 & 0.4361 & 6.7322  & 1.0419 & 0.1831 & $-2.3139$ \\
			10  & 7.8167  & 7.6710  & 0.1457 & 0.0186 & 0.3626 & 8.7442  & 0.9275 & 0.1187 & $-2.3874$ \\
			20  & 11.3979 & 11.3015 & 0.0964 & 0.0085 & 0.2912 & 12.1959 & 0.7980 & 0.0700 & $-2.4588$ \\
			50  & 19.8514 & 19.7981 & 0.0533 & 0.0027 & 0.2130 & 20.4798 & 0.6284 & 0.0317 & $-2.5370$ \\
			100 & 30.8992 & 30.8656 & 0.0336 & 0.0011 & 0.1682 & 31.4134 & 0.5142 & 0.0166 & $-2.5818$ \\[5pt]
			\hline\hline
	\end{tabular}}
	\label{parameters_funnel}
\end{table}

\section{Langer Approach {\it vs} Exact B-S Quantization Rule (\ref{radial-L})}

%\textcolor{red}{I included this Section for completeness. In my opinion, a comparison with "standard" approaches was needed. }
\label{vsLanger}

Let us compare  the accuracy of the B-S quantization rule (\ref{radial-L}) with the one emerging 
from the well-known quantization rule based on the introduction of the Langer correction \cite{Langer:1937} 
into the original potential in the standard B-S quantization rule. At $d=3$, this Bohr-Sommerfeld quantization 
rule with Langer correction inserted reads\footnote{For the three-dimensional case, the Langer correction 
corresponds to the replacement  $\ell(\ell+1)\rar(\ell+\frac{1}{2})^2$ in the centrifugal potential.}
	\begin{equation}
	\int_{r_A}^{r_B}\sqrt{E_{BS}^{(L)}-V(r)-\frac{(\ell+\frac{1}{2})^2}{r^2}}\,dr\ =\
	\pi\left(n_r\ +\ \frac{1}{2}\right)\ ,\qquad n_r,\ell \ =\  0,1,2, ...\ .
	\label{radialLanger}
\end{equation}
where we set $\hbar=1$.	
The superscript $(L)$ stands for the energy with  the Langer correction is taken into account, $E_{BS}^{(L)}$. 
This was introduced to distinguish from $E_{BS}$ already used in (\ref{radial-L}). For both the Coulomb and 
the harmonic oscillator potentials, the quantization rule (\ref{radialLanger}) leads to the exact energy spectrum.~\footnote{We suspect that this, together with the correct asymptotic behavior of the WKB wave 
function near $r=0$, motivated the specific form of the correction.}

We performed a numeric analysis for the quartic oscillator potential $V(r)=r^4$. In Table~\ref{Comparison} 
the energies coming from the quantization rules (\ref{radial-L}) and (\ref{radialLanger}) are compared with 
the exact spectrum found in the Lagrange Mesh Method for $\ell=0,1,5$ and different $n_r=0,5,10,20,100$.  
From those results, we observe that the energy $E_{BS}^{(L)}$  systematically provides a lower bound for 
the exact energy. As can be seen, in general, $E_{BS}^{(L)}$ and $E_{BS}$ are sufficiently close to 
the exact energy $E_{exact}$. For the $S$-states $E_{BS}$ is closer to $E_{exact}$ than $E_{BS}^{(L)}$ 
while for $\ell > 0$ the situation is reversed: $E_{BS}^{(L)}$ is closer to $E_{exact}$ than $E_{BS}$.

\begin{table}[h]
	\caption{Quartic Oscillator: comparison between the quantization rules  (\ref{radialLanger}),
     (\ref{radial-L}) and the exact energies at $\ell=0,1,5$ and different radial quantum numbers $n_r$.}
\label{Comparison}
	{\setlength{\tabcolsep}{0.4cm}	
\begin{tabular}{ccccc}
\hline \hline
\rule{-2pt}{4ex}		
		$\ell$&	$n_r$&	$E_{exact}$ & $E_{BS}^{(L)}$ (\ref{radialLanger})  & $E_{BS}$   (\ref{radial-L})  
\\[5pt]
		\hline
\rule{-2pt}{4ex}	
   & 0     &3.7997         & 3.6933    &  3.7519   \\
   & 5     &56.7342        & 56.7039   &  56.7191  \\
  0 & 10    &130.6421       & 130.6221  & 130.6321  \\
   & 20    &313.9580       & 313.9451  & 313.9516  \\
   & 100   &2581.2634      & 2581.2589 & 2581.2612  
\\[5pt]
\hline \hline
\rule{-2pt}{4ex}	
& 0    & 7.1084         & 7.04314      & 7.4140      \\
& 5     & 63.2881       & 63.2597      & 63.3887     \\
1 & 10 & 138.7275 & 138.7082 & 138.7955 \\
& 20    & 324.0342      & 324.02157    & 324.0787    \\
& 100   & 2598.3399     & 2598.3354    & 2598.3555    
\\[5pt]
\hline\hline
		\rule{-2pt}{4ex}		
  &	0     & 23.9406        & 23.9117    & 26.5063    \\
  &	5     & 90.3738        & 90.3516    & 91.7861    \\
5 &	10    & 171.6013       & 171.5843   & 172.6340   \\
  &	20    & 364.6617       & 364.6499   & 365.3725   \\
  &	100   & 2666.7507      & 2666.7463  & 2667.0140      
\\[5pt]
\hline \hline
\end{tabular}}
\end{table}

\section*{Conclusions}

A new Bohr-Sommerfeld type quantization rule for the $d$-dimensional radial potentials (\ref{radial-L})
is proposed for the case of $\ell$-states - the states with arbitrary
angular momentum.

The four basic requirements behind this quantization condition are fulfilled:

(i) it should coincide with the one-dimensional standard WKB quantization condition at $d=1$ 
(and at $\ell=0$), see \cite{BS-1:2021};

(ii) the B-S spectrum of the $d$-dimensional harmonic oscillator should be exact. Therefore,  
it should coincide with the $d$-dimensional WKB quantization condition for the $S$-states 
at $\ell=0$, proposed in \cite{BS-d:2023};

(iii) at large radial quantum numbers $n_r$ the B-S spectrum should approach  the exact one.

(iv) For singular potentials, condition (ii) should be replaced
by the requirement that the B-S spectrum of the $d$-dimensional
Coulomb problem be exact.

Explicit calculations for radial power-like potentials, including  the Coulomb and logarithmic potentials, 
prove the correctness of the proposal. In particular, at $d=2$, the original B-S quantization rule
reproduces the exact Coulomb energies for the $S$ states
($\ell=0$).  {In order to obtain the eigenvalues accurately, the radial Schr\"odinger equation 
(\ref{radial_se-new}) was solved using the Lagrange Mesh method with 100, 500, 1000 mesh points. 
This allows us to check for convergence and to reach the extremely high accuracies. In particular, 
for the linear potential with 100 (optimized) mesh points  the energies of the first 37 low-lying 
$S$ states can be established with at least 4 exact d.d. while with 500 optimized mesh points, 
the energies of the first 100 low-lying $S$ states can be established with exact 140 d.d., at least.}

The notion of the WKB correction, introduced for the $d=1$ case \cite{BS-1:2021} and then extended 
to the $S$ states of the $d$-dimensional radial case in \cite{BS-d:2023}, is explored. For power-like 
radial potentials, $V(r)=r^m$ with fixed $m > 0$ and quantum number $n_r$, the WKB correction $\gamma$, 
see (\ref{radial_modified}), is small for all $\ell=0,1,2,3,4,5$.
%%%%
%Checked
For fixed $d$, $m>0$ and small $\ell$, the magnitude of the WKB
correction is a monotonically decreasing function of $n_r$.
%%%%%
For fixed $d$, $n_r$ and small $\ell$, the WKB correction varies smoothly  as $m>2$ grows. 
In general, the WKB correction $\gamma$ is a smooth, dimensionless function of the parameters 
$(\la; d, m, n_r)$: for fixed $(\la; d, m)$, where $\la= \bigg(\frac{\hbar^2}{2\mu}\bigg)^{1/4} g$\ ,
see (\ref{lambda}), this function can easily be fitted by the ratio of a polynomial to the square-root 
of another polynomial \footnote{Sometimes, the square-root of a polynomial of even degree can be reduced 
to a polynomial} of the argument $n_r$. In all studied cases at $m > 0$ and $\ell=1,2,3,4,5$ 
the WKB correction $$|\gamma_+| < 1/2\ .$$

Replacing the condition (ii): {\it the B-S spectrum of the $d$-dimensional harmonic oscillator should be exact} 
by different one: {\it the B-S spectrum of the $d$-dimensional Coulomb potential should be exact}, it leads 
to a modified r.h.s. of the B-S quantization condition,
\[
    \left( n_r\ +\ \frac{d+2\ell}{4} \right)\ \rar \ \left( n_r\ +\ \frac{d+2\ell-1}{2} \right)
\]
see (\ref{radial_modified}), (\ref{radial-L-mod-nega}). Both expressions can be unified 
into a single function,
\[
    K \equiv n_r + \left( \frac{3d+6\ell-2}{8}\right) - \mbox{sign}(m) \left(\frac{d+2\ell-2}{8}\right)
\]
where the signum function $\mbox{sign}(2)=1$ (the harmonic oscillator case) and $\mbox{sign}(-1)=-1$ 
(the Coulomb potential case). It results into a natural quantization condition
\begin{equation}
\label{WKB-radial-final}
	\frac{1}{\hbar}\int_{r_A}^{r_B}\sqrt{E-V(r)}\,dr\ =\
     \pi\left(K + \gamma \right)\ ,\qquad n_r \ =\  0,1,2, ...\ ,
\end{equation}
which coincides with (\ref{radial_modified}) for non-singular at $r=0$ potentials, but including 
the logarithmic potential, with behavior $V \rar r^m$ at $r \rar 0$ for $m \geq 0$, 
when $\mbox{sign}(m)=1$. In turn, for singular potentials with $m < 0$, it coincides with 
(\ref{radial-L-mod-nega}), when $\mbox{sign}(m)=-1$. We conjecture that for both cases 
of non-singular and singular potentials, %at $\ell=0$, 
the WKB correction $$|\gamma_{\pm}| \leq 1/2\ .$$ In this article the conjecture was confirmed for 
some selected power-like potentials~\footnote{The conjecture does not hold for the funnel-type potential (\ref{funnel-potential}).}.

\begin{figure}[h]
	\centering
	\includegraphics[width=\textwidth]{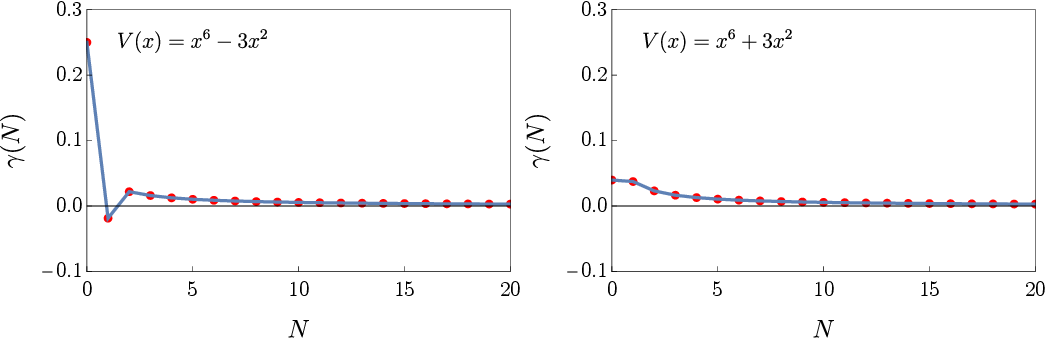}
	\caption{WKB correction for QES potential $x^6 - 3 x^2$ and its SUSY partner.}
	\label{SUSY}
\end{figure}

This article concludes the studies of the power-like potentials using the (exact) Bohr-Sommerfeld 
quantization condition (\ref{radial-L}). A natural question to ask is what would happen 
if the harmonic oscillator potential is added to the power-like potential.
It was shown in \cite{BS-d:2023} that for the $d$-dimensional quartic radial anharmonic oscillator,
\[
    V\ =\ r^2 + g^2 r^4
\]
for $d=1,2,3,6$ and $S$ states the WKB correction  $\gamma_g$ is very small for any $g^2 \geq 0$ 
and always smaller than $\gamma_{4}$ for the corresponding quartic potential,
\[
    |\gamma_g| < |\gamma_{4}|\ .
\]
Evidently, this property will remain for $\ell$-states with $\ell > 0$. Taking another example 
at $d=1$ of sextic SUSY partners,
\[
      V_{\mp}\ =\ x^6\ \pm 3 x^2\ ,
\]
with superpotential $w=x^3$, one can see that the WKB correction $\gamma$ remains very small for
both the bosonic potential $V_+$ and the fermionic potential $V_-$, see Fig. \ref{SUSY}.
Note that $V_+=x^6-3x^2$ is quasi-exactly-solvable: its ground state (at $N=0$) is known exactly
$\Psi_0=e^{-x^4/4}, E_0=0$. Its WKB correction can be found analytically, $\gamma_{V_+} (N=0)=1/4$: 
it is larger than $\gamma_{V=r^6} (N=0) \sim 0.15$. In turn,
$\gamma_{V_-} (N=0)\sim 0.04$ and it is smaller than $\gamma_{V=r^6} (N=0)$.

In a recent development of the theory of potential curves of diatomic molecules (dimers) see e.g. 
\cite{TO:2022} a new class of rational potentials was proposed,
\begin{equation}
\label{dimer}
  V\ =\ -\frac{1}{r}\ \frac{P_n(r)}{Q_{n+5}(r)}\ ,
\end{equation}
which are defined in 3-dimensional space, for neutral molecules. Here $P_n$ and $Q_{n+5}>0$ at $r > 0$ 
are polynomials of degrees $n$ and $(n+5)$, respectively. The physical meaning of these
potentials is related to a modelling of the screening of the Coulomb interaction of nuclei
by the electronic cloud. The value of the WKB correction is of great importance.
It will be studied elsewhere.

\section*{Acknowledgments}

J.C.dV. acknowledges partial financial support from the SONATA BIS 10 Grant No. 2019/34/E/ST1/00390 (Poland) 
during the initial stages of this work and from UNAM through PAPIIT project IA100926.
A.V.T. was partially supported by CONACyT Grant A1-S-17364 and DGAPA grants IN113022 
(in early stage of work), and IN104125 (Mexico). A.V.T. expresses his gratitude to I.H.E.S. (France), 
where this work was completed. 

\pagebreak

\noindent
%\bibliography{references.bib}

%%------------------ TABLE IV --------------------------

\appendix
\section{Numerical Results}
In this Appendix  selected  numerical results supporting the conclusions of the work are presented.
\begin{table}[h] 
	\caption{Energies for $n_r=0, 5, 10, 20$ for Coulomb ($m=-1$), logarithmic ($m=0$), linear ($m=1$), cubic ($m=3$), quartic ($m=4$), and sextic ($m=6$) potentials at $d=2$ and $\ell=1$. Except for Coulomb potential, all printed digits in energies rounded to four decimal digits, while in A.D., R.D. and $\gamma_+$ rounded to two significant digits .}
	{\setlength{\tabcolsep}{0.4cm}		
		\begin{tabular}{|ccccccc|}
			\hline\hline
			\multicolumn{7}{|c|}{				\rule{0pt}{4ex}	$\ell=1$}
			\\			
			$m$&$n_r$  & $E_{exact}$ & $E_{BS}$ & A.D. & R.D. & $\gamma_+$ \\[5pt]
			\hline
			\rule{0pt}{4ex}	
			\multirow{4}{*}{$-$1}
			&0  & $-$1/9           &$-$1/4         & $1.4\times 10^{-1}$   & 1.25      & 1/2    \\
			&5  & $-$1/169         &$-$1/144       & $1.0\times 10^{-3}$   & $1.7\times 10^{-1}$       &1/2          \\
			&10 &$-$1/529           &$-$1/484          &$1.8\times 10^{-4}$     &$9.3\times 10^{-2}$       &1/2          \\
			&20 &$-$1/1849           &$-$1/1764          &$2.6\times 10^{-5}$      &$4.8\times 10^{-2}$       &1/2         \\[10pt]
			%%%%%%%%%%%%%%%%%%%%%%%%%%%%%%%%%%%%%%%%%%%%%%%%%%%%%		
			\multirow{4}{*}{0}
			&0  &1.3862           &1.2655          &$1.2\times 10^{-1}$    & $8.7\times 10^{-2}$      & $1.3\times 10^{-1}$         \\
			&5  & 3.0697           &3.0573         &$1.2\times 10^{-2}$     &$4.0\times 10^{-3}$     &$7.5\times 10^{-2}$          \\
			&10 &3.6691           &3.6634          &$5.7\times 10^{-3}$    &$1.6\times 10^{-3}$    & $6.3\times 10^{-2}$         \\
			&20 &4.3126          &4.3100          &$2.6\times 10^{-3}$     &$5.9\times 10^{-4}$       & $5.4\times 10^{-2}$        \\[10pt]				
			%%%%%%%%%%%%%%%%%%%%%%%%%%%%%%%%%%%%%%%%%%%%%%%%%%%%%				
			\multirow{4}{*}{1}
			&0  &2.8721           &2.8108          &$6.1\times 10^{-2}$    & $2.1\times 10^{-2}$      & $3.3\times10^{-2}$            \\
			&5  &9.2907            & 9.2810        &$9.65\times 10^{-3}$     &$1.0\times 10^{-3}$     &$9.4\times10^{-3}$          \\
			&10 &13.9073           &13.9024          &$4.9\times 10^{-3}$    &$3.5\times 10^{-4}$    &$5.8\times10^{-3}$          \\
			&20 &21.3971          &21.3947          &$2.3\times 10^{-3}$     &$1.1\times 10^{-4}$       &$3.4\times10^{-3}$         \\[10pt]	
			%%%%%%%%%%%%%%%%%%%%%%%%%%%%%%%%%%%%%%%%%%%%%%%%%%%%%		
			\multirow{4}{*}{3}
			&0  &4.7899           &4.8600          &$7.0\times 10^{-2}$    & $1.5\times 10^{-2}$      &$-1.2\times10^{-2}$          \\
			&5  &41.7117            &41.7267         &$1.5\times 10^{-2}$     &$3.6\times 10^{-4}$     &$-1.8\times10^{-3}$          \\
			&10 &86.3491           &86.3583          &$9.2\times 10^{-3}$    &$1.1\times 10^{-4}$    &$-1.0\times10^{-4}$          \\
			&20 &187.6219          &187.6274          &$5.5\times 10^{-3}$     &$2.9\times 10^{-5}$  &$-5.1\times10^{-4}$         \\[10pt]	
			%%%%%%%%%%%%%%%%%%%%%%%%%%%%%%%%%%%%%%%%%%%%%%%%%%%%%					
			\multirow{4}{*}{4}
			&0  &5.3942         &5.5060          &$1.1\times 10^{-1}$    & $2.1\times 10^{-2}$      &$-1.5\times10^{-2}$          \\
			&5  &60.0011           &60.0307     &$3.0\times 10^{-2}$   &$5.0\times 10^{-4}$     &$-2.2\times10^{-3}$           \\
			&10 &134.6786        &134.6983   &$2.0\times 10^{-2}$    &$1.5\times 10^{-4}$    & $-1.2\times10^{-3}$          \\
			&20 &318.9923         & 319.0050 &$1.3\times 10^{-2}$     &$4.0\times 10^{-5}$   &  $-6.3\times10^{-4}$         \\[10pt]	
			%%%%%%%%%%%%%%%%%%%%%%%%%%%%%%%%%%%%%%%%%%%%%%%%%%%%%				
			\multirow{4}{*}{6}
			&0  &6.2985           &6.4066          &$1.1\times 10^{-1}$    & $1.7\times 10^{-2}$      &$-1.1\times10^{-2}$            \\
			&5  &94.1129           &94.1580         &$4.5\times 10^{-2}$     &$4.8\times 10^{-4}$     &$-1.9\times10^{-3}$            \\
			&10 &233.6994        &233.7326          &$3.3\times 10^{-2}$    &$1.4\times 10^{-4}$    &$-1.0\times10^{-3}$            \\
			&20 &616.5131         &616.5371          &$2.4\times 10^{-2}$     &$3.9\times 10^{-5}$       &$-5.5\times10^{-4}$           \\[10pt]
			\hline\hline	 	
	\end{tabular}}
	\label{firstTable-IV}
\end{table}

%%------------------ TABLE V --------------------------	

\begin{table}[h]
	\caption{Energies for $n_r=0, 5, 10, 20$ for Coulomb ($m=-1$), logarithmic ($m=0$), linear ($m=1$),
		cubic ($m=3$), quartic ($m=4$), and sextic ($m=6$) potentials at $d=2$ and $\ell=2$. Except for Coulomb potential, all printed digits in energies rounded to four decimal digits, while in A.D., R.D. and $\gamma_+$ rounded to two significant digits .}
	{\setlength{\tabcolsep}{0.4cm}		
		\begin{tabular}{|ccccccc|}
			\hline\hline
			\multicolumn{7}{|c|}{	\rule{0pt}{4ex}	$\ell=2$}
			\\			
			$m$&$n_r$  & $E_{exact}$ & $E_{BS}$ & A.D. & R.D. & $\gamma_+$ \\[5pt]
			\hline
			\rule{0pt}{4ex}	
			\multirow{4}{*}{$-$1}
			&0  & $-$1/25          &   $-$1/9       &$7.1\times 10^{-2}$    & 1.8     &       1   \\
			&5  &   $-$1/225        &$-$1/169          &$1.5\times 10^{-3}$      &$3.3\times 10^{-1}$       &   1       \\
			&10 &   $-$1/625        & $-$1/529         &$2.9\times 10^{-4}$     &$1.8\times 10^{-1}$       &   1       \\
			&20 & $-$1/2025          & $-$1/1849         &$4.7\times 10^{-5}$      &$9.5\times 10^{-2}$       &   1      \\[10pt]
			%%%%%%%%%%%%%%%%%%%%%%%%%%%%%%%%%%%%%%%%%%%%%%%%%%%%%		
			\multirow{4}{*}{0}
			&0  &1.8444           &1.6710          &$1.7\times 10^{-1}$    & $9.4\times 10^{-2}$      &$2.8\times 10^{-1}$          \\
			&5  &3.1653            &3.1373         &$2.8\times 10^{-2}$     &$8.8\times 10^{-3}$     & $1.8\times 10^{-1}$         \\
			&10 &3.7213           &3.7079          &$1.3\times 10^{-2}$    &$3.6\times 10^{-3}$    &$1.6\times 10^{-1}$          \\
			&20 & 4.3396         & 4.3336         &$6.0\times 10^{-3}$     &$1.4\times 10^{-3}$       & $1.3\times 10^{-1}$        \\[10pt]				
			%%%%%%%%%%%%%%%%%%%%%%%%%%%%%%%%%%%%%%%%%%%%%%%%%%%%%				
			\multirow{4}{*}{1}
			&0  &3.8175           & 3.6832         &$1.3\times 10^{-1}$    & $3.5\times 10^{-2}$      &$8.3\times10^{-2}$          \\
			&5  &9.8230            & 9.7897        &$3.3\times 10^{-2}$     &$3.4\times 10^{-3}$     &$3.3\times10^{-2}$          \\
			&10 &14.3387           &14.3205          &$1.8\times 10^{-2}$    &$1.3\times 10^{-3}$    &$2.2\times10^{-2}$          \\
			&20 &21.7421          &21.7330          &$9.1\times 10^{-3}$     &$4.2\times 10^{-4}$       & $2.4\times10^{-2}$        \\[10pt]	
			%%%%%%%%%%%%%%%%%%%%%%%%%%%%%%%%%%%%%%%%%%%%%%%%%%%%%		
			\multirow{4}{*}{3}
			&0  &7.6471           &7.9057          &$2.6\times 10^{-1}$          & $3.4\times 10^{-2}$      &$-4.1\times10^{-2}$          \\
			&5  &45.8482            &45.9335         &$8.2\times 10^{-2}$     &$1.9\times 10^{-3}$     &$-1.0\times10^{-2}$          \\
			&10 &91.0357           &91.0899          &$5.4\times 10^{-2}$      &$6.0\times 10^{-4}$    &$-5.7\times10^{-3}$          \\
			&20 &192.9679          &193.0008         &$3.3\times 10^{-2}$     &$1.7\times 10^{-4}$       &$-3.1\times10^{-3}$         \\[10pt]	
			%%%%%%%%%%%%%%%%%%%%%%%%%%%%%%%%%%%%%%%%%%%%%%%%%%%%%					
			\multirow{4}{*}{4}
			&0  &8.9281         &9.4542          &$5.3\times 10^{-1}$    & $5.9\times 10^{-2}$      &$-6.3\times10^{-2}$          \\
			&5  &66.5959         & 66.7918         &$2.0\times 10^{-1}$     &$2.9\times 10^{-3}$     &$-1.4\times10^{-2}$          \\
			&10 &142.7893          & 142.9231         &$1.3\times 10^{-1}$    &$9.4\times 10^{-4}$    &$-8.1\times10^{-3}$          \\
			&20 &329.0840        & 329.1722           &$8.8\times 10^{-2}$     &$2.7\times 10^{-4}$       &$-4.3\times10^{-3}$         \\[10pt]	
			%%%%%%%%%%%%%%%%%%%%%%%%%%%%%%%%%%%%%%%%%%%%%%%%%%%%%				
			\multirow{4}{*}{6}
			&0  &10.8220           &11.7697          &$9.5\times 10^{-1}$    & $8.8\times 10^{-2}$      &$-8.2\times10^{-2}$            \\
			&5  &105.7358            &106.1696         &$4.3\times 10^{-1}$     &$4.1\times 10^{-3}$     &$-1.8\times10^{-2}$          \\
			&10 &249.5230           &249.8487          &$3.3\times 10^{-1}$    &$1.3\times 10^{-3}$    &$-1.0\times10^{-2}$          \\
			&20 &638.4488          & 638.6869         &$2.4\times 10^{-1}$     &$3.7\times 10^{-4}$       &$-5.3\times10^{-3}$         \\[10pt]	
			\hline\hline	 	
	\end{tabular}}
	\label{secondTable-V}
\end{table}

%%------------------ TABLE VI --------------------------

\begin{table}[h]
	\caption{Energies for $n_r=0, 5, 10, 20$ for Coulomb ($m=-1$), logarithmic ($m=0$), linear ($m=1$),
		cubic ($m=3$), quartic ($m=4$), and sextic ($m=6$) potentials at $d=2$ and $\ell=3$. Except for Coulomb potential, all printed digits in energies rounded to four decimal digits, while in A.D., R.D. and $\gamma_+$ rounded to two significant digits .}
	{\setlength{\tabcolsep}{0.4cm}		
		\begin{tabular}{|ccccccc|}
			\hline\hline
			\multicolumn{7}{|c|}{	\rule{0pt}{4ex}	$\ell=3$}
			\\			
			$m$&$n_r$  & $E_{exact}$ & $E_{BS}$ & A.D. & R.D. & $\gamma_+$ \\[5pt]
			\hline
			\rule{0pt}{4ex}	
			\multirow{4}{*}{$-$1}
			&0  &$-$1/49           &$-$1/16          &$4.2\times 10^{-2}$    & 2.1      &    3/2      \\
			&5  & $-$1/289          &$-$1/196          &$1.6\times 10^{-3}$      &$4.7\times 10^{-1}$       &     3/2      \\
			&10 & $-$1/729          &$-$1/576          &$3.6\times 10^{-4}$     &$2.7\times 10^{-1}$       &   3/2        \\
			&20 & $-$1/2209          &$-$1/1936          &$6.4\times 10^{-5}$      &$1.4\times 10^{-1}$       &     3/2     \\[10pt]
			%%%%%%%%%%%%%%%%%%%%%%%%%%%%%%%%%%%%%%%%%%%%%%%%%%%%%		
			\multirow{4}{*}{0}
			&0  &2.1578           &1.9587          &$2.0\times 10^{-1}$    & $9.2\times 10^{-2}$      & $4.4\times 10^{-1}$         \\
			&5  &3.2545            &3.2114         &$4.3\times 10^{-2}$     &$1.3\times 10^{-2}$     &$3.1\times 10^{-1}$          \\
			&10 &3.7719           &3.7504          &$2.2\times 10^{-2}$    &$5.7\times 10^{-3}$    & $2.6\times 10^{-1}$         \\
			&20 &4.3664          &4.3566          &$9.9\times 10^{-3}$     &$2.3\times 10^{-3}$       & $2.2\times 10^{-1}$        \\[10pt]				
			%%%%%%%%%%%%%%%%%%%%%%%%%%%%%%%%%%%%%%%%%%%%%%%%%%%%%				
			\multirow{4}{*}{1}
			&0  &4.6582           & 4.4618         &$2.0\times 10^{-1}$    & $4.2\times 10^{-2}$      &$1.3\times10^{-1}$          \\
			&5  & 10.3478           &10.2855         &$6.2\times 10^{-2}$     &$6.0\times 10^{-3}$     &$6.4\times10^{-2}$          \\
			&10 &14.7685           &14.7327          &$3.6\times 10^{-2}$    &$2.4\times 10^{-3}$    &$4.4\times10^{-2}$          \\
			&20 &22.0874          & 22.0687         &$1.9\times 10^{-2}$     &$8.5\times 10^{-4}$       &$2.8\times10^{-2}$         \\[10pt]	
			%%%%%%%%%%%%%%%%%%%%%%%%%%%%%%%%%%%%%%%%%%%%%%%%%%%%%		
			\multirow{4}{*}{3}
			&0  & 10.6925          & 11.1653         &$4.7\times 10^{-1}$    & $4.4\times 10^{-2}$      & $-7.1\times10^{-2}$         \\
			&5  &50.0150            &50.2054         &$1.9\times 10^{-1}$     &$3.8\times 10^{-3}$     & $-2.2\times10^{-2}$         \\
			&10 &95.7379           &95.8629          &$1.3\times 10^{-1}$    &$1.3\times 10^{-3}$    & $-1.3\times10^{-2}$         \\
			&20 &198.3220          &198.3994          &$7.7\times 10^{-2}$     &$3.9\times 10^{-4}$       &$-7.1\times10^{-3}$         \\[10pt]	
			%%%%%%%%%%%%%%%%%%%%%%%%%%%%%%%%%%%%%%%%%%%%%%%%%%%%%					
			\multirow{4}{*}{4}
			&0  & 12.8431          &13.8743          &1.0    & $8.0\times 10^{-2}$    &    $-1.1\times 10^{-1}$         \\
			&5  &73.2766           &73.7286         &$4.5\times 10^{-1}$     &$6.2\times 10^{-3}$     & $-3.2\times 10^{-2}$          \\
			&10 &150.9522          & 151.2679         &$3.2\times 10^{-1}$    &$2.1\times 10^{-3}$    &   $-1.9\times 10^{-2}$        \\
			&20 & 339.2076         &339.4185          &$2.1\times 10^{-1}$     &$6.2\times 10^{-4}$       &   $-1.0\times 10^{-2}$       \\[10pt]	
			%%%%%%%%%%%%%%%%%%%%%%%%%%%%%%%%%%%%%%%%%%%%%%%%%%%%%				
			\multirow{4}{*}{6}
			&0  &16.0508         &18.1207            &2.1                                     & $1.3\times 10^{-1}$      &$-1.6\times 10^{-1}$                \\
			&5  &117.6050            &118.6526      &1.0                                    &$8.9\times 10^{-3}$     &$-4.1\times 10^{-2}$           \\
			&10 &265.5212          &266.3190       &$8.0\times 10^{-3}$    &$3.0\times 10^{-3}$    &$-2.4\times 10^{-2}$            \\
			&20 &660.5071        &661.0957        &$5.9\times 10^{-4}$     &$8.9\times 10^{-4}$       &$-1.3\times 10^{-2}$           \\[10pt]	
			\hline\hline	 	
	\end{tabular}}
	\label{thirdTable-VI}
\end{table}

%%------------------ TABLE VII --------------------------	

\begin{table}[h]
	\caption{Energies for $n_r=0, 5, 10, 20$ for Coulomb ($m=-1$), logarithmic ($m=0$), linear ($m=1$),
		cubic ($m=3$), quartic ($m=4$), and sextic ($m=6$) potentials at $d=2$ and $\ell=4$. Except for Coulomb potential, all printed digits in energies rounded to four decimal digits, while in A.D., R.D. and $\gamma_+$ rounded to two significant digits .}
	{\setlength{\tabcolsep}{0.4cm}		
		\begin{tabular}{|ccccccc|}
			\hline\hline
			\multicolumn{7}{|c|}{	\rule{0pt}{4ex}	$\ell=4$}
			\\			
			$m$&$n_r$  & $E_{exact}$ & $E_{BS}$ & A.D. & R.D. & $\gamma_+$ \\[5pt]
			\hline
			\rule{0pt}{4ex}	
			\multirow{4}{*}{$-$1}
			&0  &$-$1/81           &  $-$1/25        &$2.8\times 10^{-2}$    & 2.2      &      2    \\
			&5  &$-$1/361           &$-$1/225          &$1.7\times 10^{-3}$      &$6.0\times 10^{-1}$       &    2      \\
			&10 &$-$1/841           &$-$1/625          &$4.1\times 10^{-4}$     &$3.5\times 10^{-1}$       &    2      \\
			&20 &$-$1/2401           &$-$1/2025          &$7.7\times 10^{-5}$      &$1.9\times 10^{-1}$       &   2      \\[10pt]
			%%%%%%%%%%%%%%%%%%%%%%%%%%%%%%%%%%%%%%%%%%%%%%%%%%%%%		
			\multirow{4}{*}{0}
			&0  &2.3963           & 2.1818         &$2.1\times 10^{-1}$    & $9.0\times 10^{-2}$      &$5.9\times 10^{-1}$          \\
			&5  & 3.3374           &3.2804         &$5.6\times 10^{-2}$     &$1.7\times 10^{-2}$     &$4.4\times 10^{-1}$          \\
			&10 &3.8209           &3.7912          &$3.0\times 10^{-2}$    &$7.8\times 10^{-3}$    &$3.8\times 10^{-1}$          \\
			&20 &4.3929          &4.3790          &$1.4\times 10^{-2}$     &$3.2\times 10^{-3}$       &$3.2\times 10^{-1}$         \\[10pt]				
			%%%%%%%%%%%%%%%%%%%%%%%%%%%%%%%%%%%%%%%%%%%%%%%%%%%%%				
			\multirow{4}{*}{1}
			&0  &5.4290           & 5.17751         &$2.5\times 10^{-1}$    & $4.6\times 10^{-2}$      &$1.8\times10^{-1}$          \\
			&5  & 10.8639           &10.7697         &$9.4\times 10^{-2}$     &$8.7\times 10^{-3}$     & $9.9\times10^{-2}$         \\
			&10 &15.1956           &15.1391          &$5.6\times 10^{-2}$    &$3.7\times 10^{-3}$    &$7.0\times10^{-2}$          \\
			&20 &22.4322          &22.4018          &$3.0\times 10^{-2}$     &$1.4\times 10^{-3}$       &$4.6\times10^{-2}$         \\[10pt]	
			%%%%%%%%%%%%%%%%%%%%%%%%%%%%%%%%%%%%%%%%%%%%%%%%%%%%%		
			\multirow{4}{*}{3}
			&0  &13.8888           &14.5936         &$7.0\times 10^{-1}$    & $5.1\times 10^{-2}$      &  $-1.0\times10^{-1}$        \\
			&5  &54.2150            &54.5389         &$3.2\times 10^{-1}$     &$6.0\times 10^{-3}$     & $-3.7\times10^{-2}$         \\
			&10 &100.4572           &100.6758          &$2.2\times 10^{-1}$    &$2.2\times 10^{-3}$    &  $-2.3\times10^{-2}$        \\
			&20 &203.6848          &203.8225          &$1.4\times 10^{-1}$     &$6.8\times 10^{-4}$    &  $-1.3\times10^{-2}$       \\[10pt]	
			%%%%%%%%%%%%%%%%%%%%%%%%%%%%%%%%%%%%%%%%%%%%%%%%%%%%%					
			\multirow{4}{*}{4}
			&0  &17.0777      &18.6821          &1.6    &$9.4\times 10^{-2}$       & $-1.6\times 10^{-1}$          \\
			&5  &80.0467     & 80.8327        &$7.9\times 10^{-1}$     &$9.8\times 10^{-3}$     &  $-5.5\times 10^{-2}$         \\
			&10 &159.1696   &159.7295          &$5.6\times 10^{-1}$    &$3.5\times 10^{-3}$    &  $-3.3\times 10^{-2}$         \\
			&20 &349.3640  & 349.7426         &$3.8\times 10^{-1}$     &$1.1\times 10^{-3}$       &  $-1.8\times 10^{-2}$        \\[10pt]	
			%%%%%%%%%%%%%%%%%%%%%%%%%%%%%%%%%%%%%%%%%%%%%%%%%%%%%				
			\multirow{4}{*}{6}
			&0  &21.9019          &25.3245         &3.4    & $1.6\times 10^{-1}$      &  $-2.3\times 10^{-1}$         \\
			&5  &129.7226         &131.5898        &1.9    &$1.4\times 10^{-2}$     &  $-7.1\times 10^{-2}$        \\
			&10 &281.6964        &283.1362         &1.4   &$5.1\times 10^{-3}$    & $-4.2\times 10^{-2}$         \\
			&20 &682.6894        &683.7607          &1.1   &$1.6\times 10^{-3}$       &$-2.4\times 10^{-2}$         \\[10pt]	
			\hline\hline	 	
	\end{tabular}}
	\label{fourthTable-VII}
\end{table}

%%------------------ TABLE VIII --------------------------	

\begin{table}[h]
	\caption{Energies for $n_r=0, 5, 10, 20$ for Coulomb ($m=-1$), logarithmic ($m=0$), linear ($m=1$),
		cubic ($m=3$), quartic ($m=4$), and sextic ($m=6$) potentials at $d=2$ and $\ell=5$. Except for Coulomb potential, all printed digits in energies rounded to four decimal digits, while in A.D., R.D. and $\gamma_+$ rounded to two significant digits .}
	{\setlength{\tabcolsep}{0.4cm}		
		\begin{tabular}{|ccccccc|}
			\hline\hline
			\multicolumn{7}{|c|}{	\rule{0pt}{4ex}	$\ell=5$}
			\\			
			$m$&$n_r$  & $E_{exact}$ & $E_{BS}$ & A.D. & R.D. & $\gamma_+$ \\[5pt]
			\hline
			\rule{0pt}{4ex}	
			\multirow{4}{*}{$-$1}
			&0  &$-$1/121           &$-$1/36          &$2.0\times 10^{-2}$    &2.4     &  5/2        \\
			&5  &$-$1/441           &$-$1/256          &$1.6\times 10^{-3}$      &$7.2\times 10^{-1}$       & 5/2         \\
			&10 &$-$1/961           &$-$1/676          &$4.4\times 10^{-4}$     &$4.2\times 10^{-1}$       &   5/2       \\
			&20 &$-$1/2601           &$-$1/2116          &$8.8\times 10^{-5}$      &$2.3\times 10^{-1}$       &  5/2       \\[10pt]
			%%%%%%%%%%%%%%%%%%%%%%%%%%%%%%%%%%%%%%%%%%%%%%%%%%%%%		
			\multirow{4}{*}{0}
			&0  & 2.5887          &2.3641          &$2.2\times 10^{-1}$    & $8.7\times 10^{-2}$      &$7.6\times 10^{-1}$          \\
			&5  & 3.4146           & 3.3450        &$7.0\times 10^{-2}$     &$2.0\times 10^{-2}$     &$5.8\times 10^{-1}$          \\
			&10 &3.8680           & 3.8305         &$3.8\times 10^{-2}$    &$9.7\times 10^{-3}$    &$5.0\times 10^{-1}$          \\
			&20 &4.4190          &4.4010          &$1.8\times 10^{-2}$     &$4.1\times 10^{-3}$       &$4.2\times 10^{-1}$         \\[10pt]				
			%%%%%%%%%%%%%%%%%%%%%%%%%%%%%%%%%%%%%%%%%%%%%%%%%%%%%				
			\multirow{4}{*}{1}
			&0  &6.1487           &5.8467          &$3.0\times 10^{-1}$    & $4.9\times 10^{-2}$      &$2.6\times10^{-1}$             \\
			&5  &11.3709            &11.2431         &$1.3\times 10^{-1}$     &$1.1\times 10^{-2}$     &$1.4\times10^{-1}$             \\
			&10 &15.6195           &15.5402          &$7.9\times 10^{-2}$    &$5.1\times 10^{-3}$    &$1.0\times10^{-1}$             \\
			&20 &22.7763          &22.7324          &$4.4\times 10^{-2}$     &$1.9\times 10^{-3}$       &$6.7\times10^{-2}$            \\[10pt]	
			%%%%%%%%%%%%%%%%%%%%%%%%%%%%%%%%%%%%%%%%%%%%%%%%%%%%%		
			\multirow{4}{*}{3}
			&0  &17.2123           &18.1626          &$9.5\times 10^{-1}$    & $5.5\times 10^{-2}$      &$-1.3\times10^{-1}$          \\
			&5  &58.4497            & 58.9306        &$4.8\times 10^{-1}$     &$8.2\times 10^{-3}$     &$-5.4\times10^{-2}$          \\
			&10 &105.1948           &105.5274          &$3.2\times 10^{-1}$    &$3.2\times 10^{-3}$    &$-3.4\times10^{-2}$          \\
			&20 &209.0568          &209.2697          &$2.1\times 10^{-1}$     &$1.0\times 10^{-3}$       &$-2.0\times10^{-2}$         \\[10pt]	
			%%%%%%%%%%%%%%%%%%%%%%%%%%%%%%%%%%%%%%%%%%%%%%%%%%%%%					
			\multirow{4}{*}{4}
			&0  &21.5900        &23.8232          &2.2                             & $1.0\times 10^{-1}$      & $-2.1\times 10^{-1}$         \\
			&5  &86.9084           &88.09645         &1.2                          &$1.4\times 10^{-2}$     & $-8.1\times 10^{-2}$         \\
			&10 &167.4431         &168.3047          &$8.6\times 10^{-1}$    &$5.1\times 10^{-3}$    &  $-5.0\times 10^{-2}$        \\
			&20 &359.5539        &360.1436          &$5.9\times 10^{-1}$     &$1.6\times 10^{-3}$       & $-2.8\times 10^{-2}$        \\[10pt]	
			%%%%%%%%%%%%%%%%%%%%%%%%%%%%%%%%%%%%%%%%%%%%%%%%%%%%%				
			\multirow{4}{*}{6}
			&0  &28.3158           &33.2899          &5.0  & $1.8\times 10^{-1}$      & $-3.1\times 10^{-1}$         \\
			&5  &142.0888         &144.9657         &2.9    &$2.0\times 10^{-2}$     &$-1.1\times 10^{-1}$          \\
			&10 &298.0500        &300.2931          &2.2    &$7.5\times 10^{-3}$    & $-6.5\times 10^{-2}$         \\
			&20 &704.9967        &706.6788          &1.7    &$2.4\times 10^{-3}$       & $-3.7\times 10^{-2}$         \\[10pt]	
			\hline\hline	 	
	\end{tabular}}
	\label{fifthTable-VIII}
	%	\label{finald2}
\end{table}

%%------------------ TABLE IX --------------------------	

\begin{table}[h]
	\caption{Energies for $n_r=0, 5, 10, 20$ for Coulomb ($m=-1$), logarithmic ($m=0$), linear ($m=1$),
		cubic ($m=3$), quartic ($m=4$), and sextic ($m=6$) potentials at $d=3$ and $\ell=1$. Except for Coulomb potential, all printed digits in energies rounded to four decimal digits, while in A.D., R.D. and $\gamma_+$ rounded to two significant digits .}
	{\setlength{\tabcolsep}{0.4cm}		
		\begin{tabular}{|ccccccc|}
			\hline\hline
			\multicolumn{7}{|c|}{	\rule{0pt}{4ex}	$\ell=1$}
			\\			
			$m$&$n_r$  & $E_{exact}$ & $E_{BS}$ & A.D. & R.D. & $\gamma_+$ \\[5pt]
			\hline
			\rule{0pt}{4ex}	
			\multirow{4}{*}{$-$1}
			&0  &$-$1/16           &$-$4/25          &$9.8\times 10^{-2}$    & 1.6      &  3/4        \\
			&5  &$-$1/196           &$-$4/625          &$1.3\times 10^{-3}$      &$2.5\times 10^{-1}$       & 3/4         \\
			&10 &$-$1/576           &$-$4/2025          &$2.4\times 10^{-4}$     &$1.4\times 10^{-1}$       & 3/4         \\
			&20 &$-$1/1936           &$-$4/7225          &$3.7\times 10^{-5}$      &$7.2\times 10^{-2}$       &  3/4       \\[10pt]
			%%%%%%%%%%%%%%%%%%%%%%%%%%%%%%%%%%%%%%%%%%%%%%%%%%%%%		
			\multirow{4}{*}{0}
			&0  &1.6411           &1.4887          &1.5$\times 10^{-1}$    & 9.3$\times 10^{-2}$      &$2.1\times10^{-1}$          \\
			&5  &3.1183           &3.0981          &2.0$\times 10^{-2}$      &6.5$\times 10^{-3}$       & $1.3\times10^{-1}$         \\
			&10 &3.6954           &3.6859          &9.5$\times 10^{-3}$     &2.6$\times 10^{-3}$       & $1.1\times10^{-1}$         \\
			&20 &4.3261           &4.3219          &4.2$\times 10^{-3}$      &9.8$\times 10^{-4}$       &$9.0\times10^{-2}$         \\[10pt]				
			%%%%%%%%%%%%%%%%%%%%%%%%%%%%%%%%%%%%%%%%%%%%%%%%%%%%%				
			\multirow{4}{*}{1}
			&0  &3.3613           &3.2616          &1.0$\times 10^{-1}$    & 3.0$\times 10^{-2}$      &$5.8\times10^{-2}$          \\
			&5  &9.5576           &9.5371        &2.1$\times 10^{-2}$      &2.2$\times 10^{-3}$       &$2.0\times10^{-2}  $        \\
			&10 &14.1231          &14.1122          &1.1$\times 10^{-2}$     &7.7$\times 10^{-4}$       &$1.3\times10^{-2}$          \\
			&20 &21.5695         &21.5642          &5.3$\times 10^{-3}$      &2.5$\times 10^{-5}$       &$7.9\times10^{-3}$         \\[10pt]
			%%%%%%%%%%%%%%%%%%%%%%%%%%%%%%%%%%%%%%%%%%%%%%%%%%%%%		
			\multirow{4}{*}{3}
			&0  &6.1918           &6.3522          &1.6$\times 10^{-1}$    & 2.6$\times 10^{-2}$      &$-2.6\times10^{-2}$          \\
			&5  &43.7764           &43.8217          &4.5$\times 10^{-2}$      &1.0$\times 10^{-3}$       &$-5.4\times10^{-3}$          \\
			&10 &88.6906           &88.7189          &2.8$\times 10^{-2}$     &3.2$\times 10^{-4}$       & $-3.0\times10^{-3}$         \\
			&20 &190.2939           &190.3109          &1.7$\times 10^{-2}$      &9.0$\times 10^{-5}$       & $-1.6\times10^{-3}$        \\[10pt]
			%%%%%%%%%%%%%%%%%%%%%%%%%%%%%%%%%%%%%%%%%%%%%%%%%%%%%					
			\multirow{4}{*}{4}
			&0  &7.1084           &7.4140          & 3.1$\times 10^{-1}$    & 4.3$\times 10^{-2}$      & $-3.9\times10^{-2}$         \\
			&5  &63.2881           &63.3887          &1.0$\times 10^{-1}$      &1.6$\times 10^{-3}$       &$-7.4\times10^{-3}$          \\
			&	10 &138.7275           &138.7955          &6.8$\times 10^{-2}$     &4.9$\times 10^{-4}$       &$-4.1\times10^{-3}$          \\
			&	20 &324.0342        &324.0787          &4.4$\times 10^{-2}$      &1.4$\times 10^{-4}$       &$-2.2\times10^{-3}$         \\[10pt]
			%%%%%%%%%%%%%%%%%%%%%%%%%%%%%%%%%%%%%%%%%%%%%%%%%%%%%				
			\multirow{4}{*}{6}
			&0  &8.4655            &8.9536          &4.9$\times 10^{-1}$    & 5.8$\times 10^{-2}$      & $-4.6\times10^{-2}$           \\
			&5  & 99.8938         &100.1038          &2.1$\times 10^{-1}$      &2.1$\times 10^{-3}$       &$-8.7\times10^{-3}$            \\
			&10 &  241.5895         &241.7459          &1.6$\times 10^{-1}$     &6.5$\times 10^{-4}$       &$-4.9\times10^{-3}$            \\
			&20 &627.4657           &627.5794          &1.1$\times 10^{-1}$      &1.8$\times 10^{-4}$       &$-2.6\times10^{-3}$           \\[10pt]
			\hline\hline	 	
	\end{tabular}}
	\label{firstTabled3-IX}
\end{table}

%%------------------ TABLE X --------------------------	

\begin{table}[h]
	\caption{Energies for $n_r=0, 5, 10, 20$ for Coulomb ($m=-1$), logarithmic ($m=0$), linear ($m=1$),
		cubic ($m=3$), quartic ($m=4$), and sextic ($m=6$) potentials at $d=3$ and $\ell=2$. Except for Coulomb potential, all printed digits in energies rounded to four decimal digits, while in A.D., R.D. and $\gamma_+$ rounded to two significant digits .}
	{\setlength{\tabcolsep}{0.4cm}		
		\begin{tabular}{|ccccccc|}
			\hline\hline
			\multicolumn{7}{|c|}{	\rule{0pt}{4ex}	$\ell=2$}
			\\			
			$m$&$n_r$  & $E_{exact}$ & $E_{BS}$ & A.D. & R.D. & $\gamma_+$ \\[5pt]
			\hline
			\rule{0pt}{4ex}	
			\multirow{4}{*}{$-$1}	
			&0  &$-$1/36           &$-$4/49          &$5.4\times 10^{-2}$    & 1.9      & 5/4         \\
			&5  &$-$1/256           &$-$4/729          &$1.6\times 10^{-3}$      &$4.0\times 10^{-1}$       &  5/4        \\
			&10 &$-$1/676           &$-$4/2209          &$3.3\times 10^{-4}$     &$2.2\times 10^{-1}$       & 5/4         \\
			&20 &$-$1/2116           &$-$4/7569          &$5.6\times 10^{-5}$      &$1.2\times 10^{-1}$       & 5/4        \\[10pt]
			%%%%%%%%%%%%%%%%%%%%%%%%%%%%%%%%%%%%%%%%%%%%%%%%%%%%%		
			\multirow{4}{*}{0}
			&0  &2.0133           &1.8251          &1.9$\times 10^{-1}$    & 9.3$\times 10^{-2}$      &$3.6\times10^{-1}$          \\
			&5  &3.2107           &3.1751          &3.6$\times 10^{-2}$      &1.1$\times 10^{-2}$       &$2.4\times10^{-1}$          \\
			&10 &3.7468             &3.7294        &1.7$\times 10^{-2}$     &4.7$\times 10^{-3}$       &$2.1\times10^{-1}$          \\
			&20 &4.3531              &4.3451      &7.9$\times 10^{-3}$      &1.8$\times 10^{-3}$       &$1.7\times10^{-1}$         \\[10pt]		
			%%%%%%%%%%%%%%%%%%%%%%%%%%%%%%%%%%%%%%%%%%%%%%%%%%%%%%%
			\multirow{4}{*}{1}
			&0  &4.2482           &4.0818          &1.7$\times 10^{-1}$    & 3.9$\times 10^{-2}$           &$1.1\times10^{-1}$            \\
			&5  &10.0864           &10.0391          &4.7$\times 10^{-2}$      &4.7$\times 10^{-3}$       &$4.8\times10^{-2}$            \\
			&10 &14.5539           &14.5273          &2.7$\times 10^{-2}$     &1.8$\times 10^{-3}$       &$3.2\times10^{-2}$            \\
			&20 &21.9148           &21.9012          &1.4$\times 10^{-2}$      &6.2$\times 10^{-4}$       &$2.0\times10^{-2}$           \\[10pt]
			%%%%%%%%%%%%%%%%%%%%%%%%%%%%%%%%%%%%%%%%%%%%%%%%%%%%%		
			\multirow{4}{*}{3}
			&0  &9.1490           &9.5122          &3.6$\times 10^{-1}$    & 4.0$\times 10^{-2}$      & $-5.6\times10^{-2}$         \\
			&5  &47.9276           &48.0615          &1.3$\times 10^{-1}$      &2.8$\times 10^{-3}$       & $-1.6\times10^{-2}$         \\
			&10 &93.3847           &93.4713          &8.7$\times 10^{-2}$     &9.3$\times 10^{-4}$       & $-9.1\times10^{-3}$         \\
			&20 &195.6439         &195.6970          &5.3$\times 10^{-2}$      &2.7$\times 10^{-4}$       & $-4.9\times10^{-3}$        \\[10pt]
			\rule{0pt}{4ex}
			\multirow{4}{*}{4}
			&0  &10.8424           &11.6115          & 7.7$\times 10^{-1}$    & 7.1$\times 10^{-2}$      &$-8.8\times10^{-2}$            \\
			&5  &69.9253           &70.2388          &3.1$\times 10^{-1}$      &4.5$\times 10^{-3}$       & $-2.3\times10^{-2}$           \\
			&10 &146.8641           &147.0807          &2.2$\times 10^{-1}$     &1.5$\times 10^{-3}$       & $-1.3\times10^{-2}$           \\
			&20 &334.1418           &334.2855          &1.4$\times 10^{-1}$      &4.3$\times 10^{-4}$       &$-7.0\times10^{-3}$           \\[10pt]
			%%%%%%%%%%%%%%%%%%%%%%%%%%%%%%%%%%%%%%%%%%%%%%%%%%%%%				
			\multirow{4}{*}{6}
			&0  &13.3542           &14.8316          &1.5                                 & 1.1$\times 10^{-1}$      &$-1.2\times10^{-1}$          \\
			&5  &111.6394           &112.3533          &$7.1\times 10^{-1}$      &6.4$\times 10^{-3}$       &$-2.9\times10^{-2}$          \\
			&10 &257.5001           &258.0401          &5.4$\times 10^{-1}$     &2.1$\times 10^{-3}$       &$-1.6\times10^{-2}$          \\
			&20 &649.4626           &649.8591          &4.0$\times 10^{-1}$      &6.1$\times 10^{-4}$       & $-8.8\times10^{-3}$        \\[10pt]			
			\hline\hline
	\end{tabular}}
	\label{Tabled3-X}
\end{table}

%%------------------ TABLE XI --------------------------	

\begin{table}[h]
	\caption{Energies for $n_r=0, 5, 10, 20$ for Coulomb ($m=-1$), logarithmic ($m=0$), linear ($m=1$),
		cubic ($m=3$), quartic ($m=4$), and sextic ($m=6$) potentials at $d=3$ and $\ell=3$. Except for Coulomb potential, all printed digits in energies rounded to four decimal digits, while in A.D., R.D. and $\gamma_+$ rounded to two significant digits .}
	{\setlength{\tabcolsep}{0.4cm}		
		\begin{tabular}{|ccccccc|}
			\hline\hline
			\multicolumn{7}{|c|}{	\rule{0pt}{4ex}	$\ell=3$}
			\\			
			$m$&$n_r$  & $E_{exact}$ & $E_{BS}$ & A.D. & R.D. & $\gamma_+$ \\[5pt]
			\hline
			\rule{0pt}{4ex}	
			\multirow{4}{*}{$-$1}
			&0  &$-$1/64           &$-$4/81          &$3.4\times 10^{-2}$    & 2.2      &7/4          \\
			&5  &$-$1/324           &$-$4/841          &$1.7\times 10^{-3}$      &$5.4\times 10^{-1}$       &7/4          \\
			&10 &$-$1/784           &$-$4/2401          &$3.9\times 10^{-4}$     &$3.1\times 10^{-1}$       &7/4          \\
			&20 &$-$1/2304           &$-$4/7921          &$7.1\times 10^{-5}$      &$1.6\times 10^{-1}$       &7/4         \\[10pt]
			%%%%%%%%%%%%%%%%%%%%%%%%%%%%%%%%%%%%%%%%%%%%%%%%%%%%%		
			\multirow{4}{*}{0}
			&0  &2.2841         &2.0764           &2.1$\times 10^{-1}$    & 9.1$\times 10^{-2}$      &  $5.2\times10^{-1}$        \\
			&5  &3.2967           &3.2465          &5.0$\times 10^{-2}$      &1.5$\times 10^{-2}$       & $3.7\times10^{-1}$         \\
			&10 &3.7966           &3.7710          &2.6$\times 10^{-2}$     &6.7$\times 10^{-2}$       &   $3.2\times10^{-1}$       \\
			&20 &4.3797          &4.3679           &1.2$\times 10^{-2}$      &2.7$\times 10^{-3}$       & $2.7\times10^{-1}$        \\[10pt]		
			%%%%%%%%%%%%%%%%%%%%%%%%%%%%%%%%%%%%%%%%%%%%%%%%%%%%%%%
			\multirow{4}{*}{1}	
			&0  &5.0509          &4.8263           &2.2$\times 10^{-1}$    & 4.4$\times 10^{-2}$      &$1.6\times10^{-1}$             \\
			&5 &10.6070         &10.5290             &7.8$\times 10^{-2}$      &7.4$\times 10^{-3}$       & $8.1\times10^{-2}$            \\
			&10 &14.9824           &14.9366          &4.6$\times 10^{-2}$     &3.1$\times 10^{-3}$       &  $5.6\times10^{-2}$           \\
			&20 &22.2598           &22.2355          &2.4$\times 10^{-2}$      &1.1$\times 10^{-3}$       &$3.7\times10^{-2}$            \\[10pt]
			%%%%%%%%%%%%%%%%%%%%%%%%%%%%%%%%%%%%%%%%%%%%%%%%%%%%%		
			\multirow{4}{*}{3}
			&0  &12.2735           &12.8603          &5.9$\times 10^{-1}$    & 4.8$\times 10^{-2}$      & $-8.6\times10^{-2}$         \\
			&5  &52.1108           &52.3647          &2.5$\times 10^{-1}$      &4.9$\times 10^{-3}$       & $-2.9\times10^{-2}$         \\
			&10 &98.0953           &98.2644          &1.7$\times 10^{-1}$     &1.7$\times 10^{-3}$       & $-1.8\times10^{-2}$         \\
			&20 &201.0023           &201.1079          &1.1$\times 10^{-1}$      &5.3$\times 10^{-4}$       & $-9.7\times10^{-3}$        \\[10pt]
			%%%%%%%%%%%%%%%%%%%%%%%%%%%%%%%%%%%%%%%%%%%%%%%%%%%%%%
			\multirow{4}{*}{4}
			&0  & 14.9234          &16.2336          & 1.3   & 8.8$\times 10^{-2}$      & $-1.4\times10^{-1}$           \\
			&5  & 76.6503          &77.2602          &6.1$\times 10^{-1}$      &8.0$\times 10^{-3}$       &$-4.3\times10^{-2}$          \\
			&10 &155.0539           &155.4843          &4.3$\times 10^{-1}$     &2.8$\times 10^{-3}$       & $-2.5\times10^{-2}$         \\
			&20 &344.2816           &344.5708          &2.9$\times 10^{-1}$      &8.4$\times 10^{-4}$       & $-1.4\times10^{-2}$        \\[10pt]
			%%%%%%%%%%%%%%%%%%%%%%%%%%%%%%%%%%%%%%%%%%%%%%%%%%%%%				
			\multirow{4}{*}{6}
			&0  &18.9027           &21.6224          &2.7                            & 1.4$\times 10^{-1}$      &$-1.9\times10^{-1}$            \\
			&5  &123.6327           &125.0654          &1.4                         &1.2$\times 10^{-2}$       &$-5.5\times10^{-2}$            \\
			&10 &273.5866           &274.6847          &1.1                        &4.0$\times 10^{-3}$       &$-3.3\times10^{-2}$            \\
			&20 &671.5827           &672.3964          &8.1$\times 10^{-1}$      &1.2$\times 10^{-3}$       & $-1.8\times10^{-2}$          \\[10pt]
			\hline\hline
	\end{tabular}}
	\label{Tabled3-XI}
\end{table}

%%------------------ TABLE XII --------------------------	

\begin{table}[h]
	\caption{Energies for $n_r=0, 5, 10, 20$ for Coulomb ($m=-1$), logarithmic ($m=0$), linear ($m=1$),
		cubic ($m=3$), quartic ($m=4$), and sextic ($m=6$) potentials at $d=3$ and $\ell=4$. Except for Coulomb potential, all printed digits in energies rounded to four decimal digits, while in A.D., R.D. and $\gamma_+$ rounded to two significant digits .}	
	{\setlength{\tabcolsep}{0.4cm}		
		\begin{tabular}{|ccccccc|}
			\hline\hline
			\multicolumn{7}{|c|}{	\rule{0pt}{4ex}	$\ell=4$}
			\\			
			$m$&$n_r$  & $E_{exact}$ & $E_{BS}$ & A.D. & R.D. & $\gamma_+$ \\[5pt]
			\hline
			\rule{0pt}{4ex}	
			\multirow{4}{*}{$-$1}	
			&0  &$-$1/100           &$-$4/121          &$2.3\times 10^{-2}$    & 2.3      &9/4          \\
			&5  &$-$1/400           &$-$4/961          &$1.7\times 10^{-3}$      &$6.6\times 10^{-1}$       &9/4          \\
			&10 &$-$1/900           &$-$4/2601          &$4.3\times 10^{-4}$     &$3.8\times 10^{-1}$       &9/4          \\
			&20 &$-$1/2500         &$-$4/8281          &$8.3\times 10^{-5}$      &$2.1\times 10^{-1}$       &9/4         \\[10pt]
			%%%%%%%%%%%%%%%%%%%%%%%%%%%%%%%%%%%%%%%%%%%%%%%%%%%%%		
			\multirow{4}{*}{0}
			&0  &2.4971           &2.2771          &2.2$\times 10^{-1}$    & 8.8$\times 10^{-2}$      & $6.8\times10^{-1}$         \\
			&5  &3.3767           &3.3132          &6.3$\times 10^{-2}$      &1.9$\times 10^{-2}$       &$5.1\times10^{-1}$          \\
			&10 &3.8447           &3.8110          &3.4$\times 10^{-2}$     &8.7$\times 10^{-3}$       &$4.4\times10^{-1}$          \\
			&20 & 4.4060          &4.3901          &1.6$\times 10^{-2}$      &3.6$\times 10^{-3}$       &$3.7\times10^{-1}$         \\[10pt]		
			%%%%%%%%%%%%%%%%%%%%%%%%%%%%%%%%%%%%%%%%%%%%%%%%%%%%%%%
			\multirow{4}{*}{1}	
			&0  &5.7944           &5.5172          &2.8$\times 10^{-1}$    &4.8$\times 10^{-2}$      & $2.1\times10^{-1}$           \\
			&5  &11.1186           &11.0077          &1.1$\times 10^{-1}$      &1.0$\times 10^{-2}$       & $1.2\times10^{-1}$           \\
			&10 &15.4080         &15.34031          &6.9$\times 10^{-2}$     &4.4$\times 10^{-3}$       &$8.4\times10^{-2}$            \\
			&20 &22.6044         &22.5674          &3.7$\times 10^{-2}$      &1.6$\times 10^{-3}$       & $5.6\times10^{-2}$          \\[10pt]
			%%%%%%%%%%%%%%%%%%%%%%%%%%%%%%%%%%%%%%%%%%%%%%%%%%%%%		
			\multirow{4}{*}{3}
			&0  &15.5358           &16.3618          &8.3$\times 10^{-1}$    & 5.3$\times 10^{-2}$      &$-1.2\times10^{-1}$          \\
			&5  &56.3279           &56.7277          &4.0$\times 10^{-1}$      &7.1$\times 10^{-3}$       &$-4.6\times10^{-2}$          \\
			&10 &102.8237          &103.0968          &2.7$\times 10^{-1}$     &2.7$\times 10^{-3}$       &$-2.8\times10^{-2}$          \\
			&20 &206.3696         &206.5431          &1.7$\times 10^{-1}$      &8.4$\times 10^{-4}$       &$-1.6\times10^{-2}$         \\[10pt]
			%%%%%%%%%%%%%%%%%%%%%%%%%%%%%%%%%%%%%%%%%%%%%%%%%%%%%	
			\multirow{4}{*}{4}	
			&0  &19.3012           &21.2137          & 1.9    & 9.9$\times 10^{-2}$      &  $-1.9\times10^{-1}$        \\
			&5  &83.4660           &84.4450          &9.8$\times 10^{-1}$      &1.2$\times 10^{-2}$       & $-6.7\times10^{-2}$        \\
			&10 &163.2992           &164.0031          &7.0$\times 10^{-1}$     &4.3$\times 10^{-3}$       &  $-4.1\times10^{-2}$        \\
			&20 &354.4547           &354.9336          &4.8$\times 10^{-1}$      &1.4$\times 10^{-3}$       &   $-2.3\times10^{-2}$       \\[10pt]
			%%%%%%%%%%%%%%%%%%%%%%%%%%%%%%%%%%%%%%%%%%%%%%%%%%%%%	
			\multirow{4}{*}{6}
			&0  &25.0416          &29.2166          & 4.2   & 1.7$\times 10^{-1}$      &$-2.7\times10^{-1}$           \\
			&5  &135.8747           &138.2238          & 2.3  &1.7$\times 10^{-2}$       &$-8.8\times10^{-2}$           \\
			&10 &289.8508           &291.6726          & 1.8 &6.3$\times 10^{-3}$       &$-5.3\times10^{-2}$           \\
			&20 &693.8274           &695.1883          &    1.4  &2.0$\times 10^{-3}$       &$-3.0\times10^{-2}$          \\[10pt]
			\hline\hline
	\end{tabular}}
	\label{Tabled3-XII}
\end{table}

%%------------------ TABLE XIII --------------------------	

\begin{table}[h]
	\caption{Energies for $n_r=0, 5, 10, 20$ for Coulomb ($m=-1$), logarithmic ($m=0$),
		linear ($m=1$), cubic ($m=3$), quartic ($m=4$), and sextic ($m=6$) potentials at $d=3$
		and $\ell=5$. Except for Coulomb potential, all printed digits in energies rounded
		to four decimal digits, while in A.D., R.D. and $\gamma_+$ rounded to
		two significant digits.}
	{\setlength{\tabcolsep}{0.4cm}		
		\begin{tabular}{|ccccccc|}
			\hline\hline
			\multicolumn{7}{|c|}{	\rule{0pt}{4ex}	$\ell=5$}
			\\			
			$m$ & $n_r$ & $E_{exact}$ & $E_{BS}$ & A.D. & R.D. & $\gamma_+$
			\\[5pt]
			\hline
			\rule{0pt}{4ex}	
			\multirow{4}{*}{$-$1}	
			&0  &$-$1/144           &$-$4/169          &$1.7\times 10^{-2}$    & 2.4     &  11/4      \\
			&5  &$-$1/484           &$-$4/1089          &$1.6\times 10^{-3}$      &$7.8\times 10^{-1}$       &11/4        \\
			&10 &$-$1/1024           &$-$4/2809          &$4.5\times 10^{-4}$     &$4.6\times 10^{-1}$       &11/4         \\
			&20 &$-$1/2704           &$-$4/8649          &$9.3\times 10^{-5}$      &$2.5\times 10^{-1}$       &11/4        \\[10pt]
			%%%%%%%%%%%%%%%%%%%%%%%%%%%%%%%%%%%%%%%%%%%%%%%%%%%%%		
			\multirow{4}{*}{0}
			&0  &2.6726           &2.4442         &2.3$\times 10^{-1}$    & 8.5$\times 10^{-2}$      & $8.3\times10^{-1}$         \\
			&5  &3.4513           &3.3132          &7.6$\times 10^{-2}$      &2.2$\times 10^{-2}$       &$6.5\times10^{-1}$          \\
			&10 &3.8910           &3.3757          &4.1$\times 10^{-2}$     &1.1$\times 10^{-2}$       &   $5.6\times10^{-1}$       \\
			&20 &4.4319           &4.4118          &2.0$\times 10^{-2}$      &4.5$\times 10^{-3}$       & $4.7\times10^{-1}$        \\[10pt]		
			%%%%%%%%%%%%%%%%%%%%%%%%%%%%%%%%%%%%%%%%%%%%%%%%%%%%%%%			
			\multirow{4}{*}{1}	
			&0  &6.4930           &6.1671          &3.3$\times 10^{-1}$    & 5.0$\times 10^{-2}$      &  $2.6\times10^{-1}$        \\
			&5  &11.6211           &11.4762          &1.4$\times 10^{-1}$      &1.2$\times 10^{-2}$       &$1.6\times10^{-1}$          \\
			&10 &15.8301           &15.7388          &9.1$\times 10^{-2}$     &5.8$\times 10^{-3}$       &$1.2\times10^{-1}$          \\
			&20 &22.9481           &22.8969          &5.1$\times 10^{-2}$      &2.2$\times 10^{-3}$       & $7.8\times10^{-2}$        \\[10pt]
			%%%%%%%%%%%%%%%%%%%%%%%%%%%%%%%%%%%%%%%%%%%%%%%%%%%%%		
			\multirow{4}{*}{3}
			&0  &18.9162           &19.9937          &1.1                               & 5.7$\times 10^{-2}$      & $-1.5\times10^{-1}$         \\
			&5  &60.5803           &61.1474          &5.7$\times 10^{-1}$      &9.4$\times 10^{-3}$       &$-6.4\times10^{-2}$          \\
			&10 &107.5708          &107.9673          &4.0$\times 10^{-1}$     &3.7$\times 10^{-3}$       & $-4.1\times10^{-2}$         \\
			&20 &211.7464           &212.0023          &2.6$\times 10^{-1}$      &$1.2\times 10^{-3}$       & $-2.3\times10^{-2}$        \\[10pt]
			%%%%%%%%%%%%%%%%%%%%%%%%%%%%%%%%%%%%%%%%%%%%%%%%%%%%%	
			\multirow{4}{*}{4}	
			&0  & 23.9406          & 26.5063      & 2.6     & 1.1$\times 10^{-1}$
			& $-2.4\times10^{-1}$          \\
			&5  & 90.3738          & 91.7861      & 1.4     & 1.6$\times 10^{-2}$
			& $-9.5\times10^{-2}$         \\
			&10 & 171.6016         & 172.6340     & 1.0     & 6.0$\times 10^{-3}$
			& $-5.9\times10^{-2}$          \\
			&20 & 364.6617         & 365.3725     & 0.71   & 1.9$\times 10^{-3}$
			& $-3.4\times10^{-2}$        \\[10pt]	
			%%%%%%%%%%%%%%%%%%%%%%%%%%%%%%%%%%%%%%%%%%%%%%%%%%%%%	
			\multirow{4}{*}{6}
			&0  &31.71928        & 37.5366       & 5.8   & 1.8$\times 10^{-1}$
			&  $-3.5\times10^{-1}$        \\
			&5  & 148.3649       & 151.8138      & 3.4   & 2.3$\times 10^{-2}$
			& $-1.3\times10^{-1}$         \\
			&10 &306.2940        & 308.9969      & 2.7   & 8.8$\times 10^{-3}$
			& $-7.7\times10^{-2}$         \\
			&20 & 716.1975       & 718.2320      & 2.0   & 2.8$\times 10^{-3}$
			& $-4.4\times10^{-2}$         \\[10pt]
			\hline\hline
	\end{tabular}}
	\label{lastTabled3-XIII}
\end{table}

\end{document}